\def\year{2025}\relax
\documentclass[letterpaper]{article} % DO NOT CHANGE THIS
\usepackage{aaai25}  % DO NOT CHANGE THIS
\usepackage{times}  % DO NOT CHANGE THIS
\usepackage{helvet}  % DO NOT CHANGE THIS
\usepackage{courier}  % DO NOT CHANGE THIS
\usepackage[hyphens]{url}  % DO NOT CHANGE THIS
\usepackage{graphicx} % DO NOT CHANGE THIS
\usepackage{natbib}  % DO NOT CHANGE THIS AND DO NOT ADD ANY OPTIONS TO IT
\usepackage{caption} % DO NOT CHANGE THIS AND DO NOT ADD ANY OPTIONS TO IT
\usepackage[dvipsnames]{xcolor}
\DeclareCaptionStyle{ruled}{labelfont=normalfont,labelsep=colon,strut=off} % DO NOT CHANGE THIS
\usepackage{algorithm}
\usepackage{algorithmic}
\usepackage{multirow}
\usepackage{graphics}
\usepackage{adjustbox}
\usepackage{xcolor}
\usepackage{amsmath}

\usepackage{newfloat}
\usepackage{listings}
\usepackage{booktabs}
\floatstyle{ruled}
\newfloat{listing}{tb}{lst}{}
\floatname{listing}{Listing}

\usepackage{amsmath}
\DeclareFontFamily{U}{skulls}{}
\DeclareFontShape{U}{skulls}{m}{n}{ <-> skull }{}

\title{Identifying and Investigating Global News Coverage of Critical Events Such as Disasters and Terrorist Attacks}

\author {
    Erica Cai\textsuperscript{\rm 1}\equalcontrib, 
    Xi Chen\textsuperscript{\rm 1}\equalcontrib, 
    Reagan Grey Keeney\textsuperscript{\rm 1}, 
    Ethan Zuckerman\textsuperscript{\rm 1}, 
    Brendan O{'}Connor\textsuperscript{\rm 1}, 
    Przemyslaw A. Grabowicz\textsuperscript{\rm 1, 2}
}
\affiliations {
    \textsuperscript{\rm 1}University of Massachusetts Amherst\\
    \textsuperscript{\rm 2} University College Dublin\\
    ecai@umass.edu, xchen4@umass.edu, reagankeeney@gmail.com,
    ethanz@umass.edu, brenocon@cs.umass.edu, grabowicz@cs.umass.edu
}

\begin{document}
\maketitle
\begin{abstract}
% Previous research hints at bias in selectivity of news coverage: \citet{eisensee2007news} find that a disaster in Africa requires a death toll that is 40 times higher than one in Eastern Europe to receive equivalent news coverage. 
% We aim to ease large-scale investigation of bias in news coverage by (1) contributing a method to efficiently extract articles about events from a database of millions of news articles, and (2) performing a large-scale investigation of bias in news coverage on disasters and conflicts.

% Selective news coverage of critical events affects public awareness about their significance, and therefore contributes to framing the agenda of public concerns. 
%While studies of news coverage are crucial, they are challenging to conduct, because methods for identifying news articles about critical events rely on expert-based heuristics that face scalability issues.
Comparative studies of news coverage are challenging to conduct because methods to identify news articles about the same event in different languages require expertise that is difficult to scale.
%
% This study presents a method, \textsc{FAME}, to spearhead large-scale investigation of biases in multilingual news coverage of critical events, where critical events are described using \textit{fingerprint}s, which is a minimal set of metadata about each event. 
We introduce an AI-powered method for identifying news articles based on an event \textsc{fingerprint}, which is a minimal set of metadata required to identify critical events.
% \ec{first discussion of fingerprint} \textsc{FAME} efficiently identifies news articles about critical events described in an event database given their fingerprints, and we subsequently perform a large-scale pilot investigation of bias in multilingual news coverage on disasters and conflicts. 
%Our event coverage identification method, \textsc{Fingerprint to Article Matching for Events (FAME)}, efficiently identifies news articles about critical events described in a database of events. Subsequently, we perform a large-scale pilot investigation of bias in multilingual news coverage of disaster and terrorist attack events. 
Our event coverage identification method, \textsc{Fingerprint to Article Matching for Events (FAME)}, efficiently identifies news articles about critical world events, specifically terrorist attacks and several types of natural disasters. 
\textsc{FAME} does not require training data and is able to automatically and efficiently identify news articles that discuss an event given its fingerprint: time, location, and class (such as storm or flood). 
% It achieves strong performance for disaster and conflict events over millions of articles collected from the Mediacloud project (cite)  compared to existing methods adapted to our task. Next, we use the method to efficiently collect articles that cover disaster and conflict events mentioned in EM-DAT, which stores natural disaster event records,  and Global Terrorism Database, which stores terrorist attack event records, during 2020. We investigate patterns of bias in news article coverage of events with respect to source country and affected country across different countries. We find that ... \ec{xi chen?}(results of regression)
The method achieves state-of-the-art performance and scales to massive databases of tens of millions of news articles and hundreds of events happening globally.
We use \textsc{FAME} to identify 27,441 articles that cover 470 natural disaster and terrorist attack events that happened in 2020. To this end, we use a massive database of news articles in three languages from MediaCloud, and three widely used, expert-curated databases of critical events: EM-DAT, USGS, and GTD. 
% Finally, we demonstrate \textsc{FAME} in a real case study and investigate patterns of bias in global news coverage of disasters and terrorist attacks over a period of six months. Our case study reveals findings that are consistent with previous literature \ec{cite}, adding external validity to the method's performance in real applications. 
Our case study reveals patterns consistent with prior literature: coverage of disasters and terrorist attacks correlates to death counts, to the GDP of a country where the event occurs, and to trade volume between the reporting country and the country where the event occurred. We share our NLP annotations and cross-country media attention data to support the efforts of researchers and media monitoring organizations.
% Particularly, we find that disasters affecting a particular country are more likely to be covered by the news media of another country if the affected country has high GDP and neighbors, shares an official language, and maintains diplomatic relations with the covering country.
%For instance, the disasters and terrorist attacks resulting in more deaths in event countries with higher GDP and larger trade volume with the reporting country are likely to achieve more news coverage. 
% disasters affecting a particular country receive more news coverage if that country has high GDP and neighbors,
\end{abstract}

\setcounter{secnumdepth}{2}

\section{Introduction}
\label{s:intro}

\begin{figure}[t]
    \centering
    \includegraphics[scale=0.29]{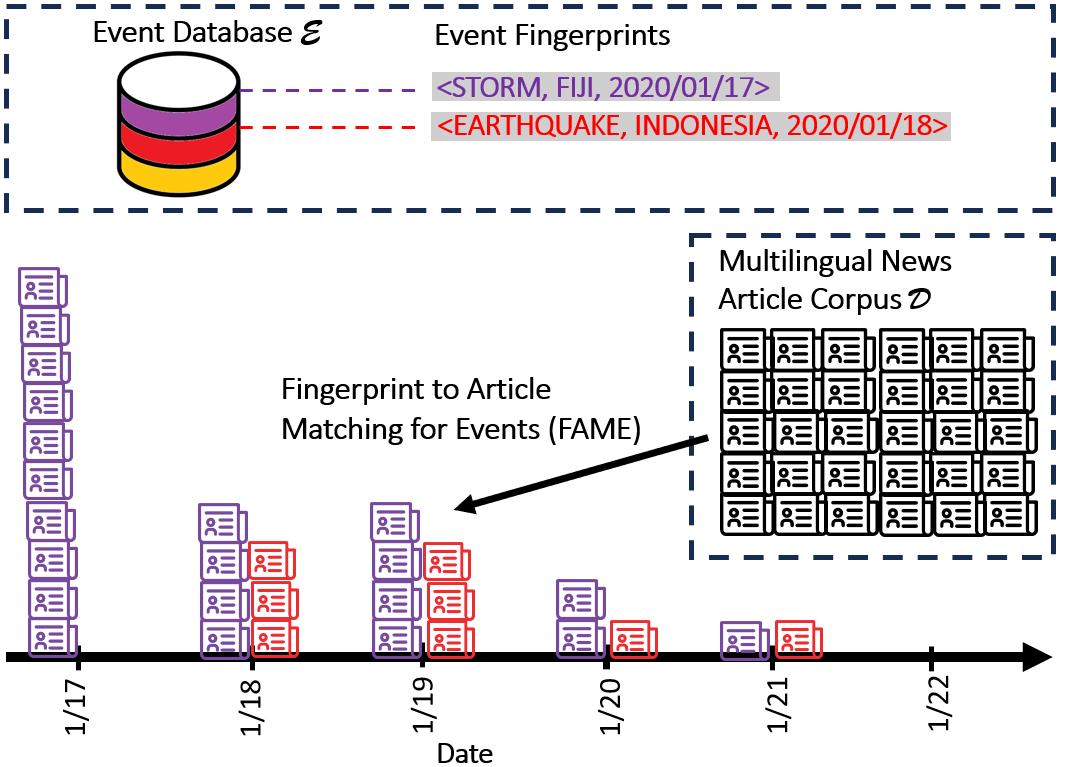}
    \caption{Our event linking method maps events in a knowledge base to articles that discuss corresponding events, at the granularity of their event fingerprints  $\langle $EventClass ($c$), 
    CountryLocation ($l$),
    Date ($t$)$\rangle$. The figure shows the real distribution over time of articles corresponding to event fingerprints $\langle$\textsc{Storm}, \textsc{Fiji}, 2020/01/17$\rangle$ and $\langle$\textsc{Earthquake}, \textsc{Indonesia}, 2020/01/18$\rangle$ (purple and red, respectively). 
    % Each article has an additional piece of metadata, the country of its publication, which we use to conduct cross-country media attention analysis.
    }\label{fig:linking}%These are examples of snippets in articles that may seem to suggest the discussion of a natural disaster due to their keywords. However, only the top article discusses one. }
\end{figure}

News organizations help shape the public's perception of disasters and terrorist attacks by choosing how, and how much, to discuss them. The coverage an event receives influences what events the public concerns itself with, a phenomenon called ``agenda setting''. 

Previous research suggests that coverage biases --- increased coverage of certain events based on the political or economic characteristics of the location in which they transpired ---  have significant effects on media attention toward natural disasters \cite{eisensee2007news}, civil war \cite{baum2015reporting}, terrorism \cite{Hellmueller02012022}, international political violence \cite{Dietrich01112020}, etc.---for example, a disaster in Africa may require a death toll 40 times higher than one in Eastern Europe to receive equivalent US\ television news coverage, and such skews in media attention even affect the amount of provided humanitarian aid\ \cite{eisensee2007news}.

Comprehensive investigation of biases in global news coverage is challenging in part  because it requires efficient, accurate identification of news articles about the same events in different languages. Manual article labeling, prevalent in social science research, is costly and difficult to accomplish on a large scale, limiting the size of corpora that can be studied \cite{Krippendorff2019content,grimmer2022text}. On the other hand, automated natural language processing (NLP) methods for cross-document event analysis often require vast amounts of labeled data for training, which is costly \cite{eirew-etal-2022-cross,eirew-etal-2021-wec}, or they have task specifications incompatible with ours --- for example, operating only on single sentences \cite{huang2018zero,zhang-etal-2022-zero}. %lack robustness in zero-shot settings

% In this paper, we introduce a novel task of identifying news articles about specific events across different languages, and a method, \textsc{FAME}, which efficiently performs the task, linking articles against events from a provided event database.
In this paper, we introduce a novel task of linking events from an event database with news articles.
Then, we contribute a novel method, \textsc{FAME}, which efficiently identifies full articles about specific events across different languages.
% , linking articles against events from a provided event database.
% , enhancing the study of news coverage biases concerning disasters and terrorist attacks. 
This method leverages a minimal set of information from a database about an event---date, country location, and event class---as a \textit{fingerprint} and accurately identifies relevant articles without requiring training data (\textbf{F}ingerprint to \textbf{A}rticle \textbf{M}atching for \textbf{E}vents).
%, a significant departure from traditional approaches that focus on sentences, quotations, titles, or headlines. 
Our multilingual approach includes two steps: matching on keywords and publication date to assemble candidate articles per event, followed by a more refined semantic filter using large language model question-answering (LLM QA) to eliminate false positives.

We rigorously evaluate \textsc{FAME}'s precision and recall for linking news articles in three languages against events of eight classes (terrorist attacks, and seven types of natural disasters) from three different databases.
Unlike conventional NLP techniques that rely on large training datasets, our method achieves superior performance, with average F1 of $94.0$ on linking events of the eight classes to English articles, without any training data by using GPT-3.5-turbo \cite{ouyang2022training}.
We collect manual annotations across dozens of events, and thousands of articles, finding a significant variation in performance between events.
% And, since we collect manual annotations across dozens of events (and thousands of articles), we can break down performance by event, finding that a single performance number can hide significant variation in performance between events.
%finding that the typical averaged evaluations from NLP hide significant performance variation at the \emph{event}-level, which is closer to the text measurement questions of interest \cite{halterman-etal-2021-corpus}.
We release data of 6,468 annotated event-news pairs to support the development of future event-to-news linking methods.\footnote{https://zenodo.org/records/14667152} To facilitate replication of this study we release its code.\footnote{https://github.com/social-info-lab/disaster\_event\_analysis}
% \footnote{URL for release at publication; for review, included in supplementary materials.}
% \bto{I drafted something, and promised supp material upload.  Erica and Xi Chen should update to be accurate}

In a pilot study applying \textsc{FAME} to a massive corpus of tens of millions of multilingual news articles, we identified news in English, Spanish, and French about 470 disaster and terrorist attack events happening globally in the first six months of 2020.
Through regression analysis, we quantify the role of various country-level factors for media coverage of disasters and terrorist attacks, spanning events in over a hundred countries.
% 124 countries as reported by news sources from 139 countries. 
We find that the number of deaths associated with an event and the economic characteristics of the affected countries correlate most strongly with media coverage, consistent with previous literature \cite{eisensee2007news, xi_global_news_2024}, underscoring the validity of this methodological approach.
In comparison to existing media studies, our research examines patterns in news coverage of a large number of critical events across 128 countries, while adding nuance to existing knowledge of biases in news coverage. For instance, our findings suggest that GDP, rather than geography, explains decreased news coverage of disasters in Africa. We anticipate future studies to build upon these results by applying \textsc{FAME} to datasets spanning much longer time periods.

\section{Related Work}
\label{s:relatedwork}

\textbf{Extracting articles discussing events.} Here, we describe the different approaches to event extraction that have been developed in the areas of NLP and social sciences. 

On the one hand, many information extraction methods from the NLP literature can enable event extraction, but most face significant challenges when deployed at scale. High performing methods for event extraction %- as well as those for the similar task of relation extraction - 
require large amounts of training data \cite{wadden2019entity,chen2020reading,du2020event,liu2020event,li2020event,lin2020joint,li2021document,ahmad2021gate,lu2021text2event}, and existing zero-shot methods tend to perform poorly and lack robustness \cite{huang2018zero,zhang-etal-2021-zero,lyu2021zero,zhang-etal-2022-zero,cai2024montecarlolanguagemodel}. Further, most existing methods aim to extract events from individual sentences, while real-world applications require extractions from paragraphs or full documents. 

A relatively small NLP literature has examined extracting events as they are discussed \emph{across} many documents, which is necessary for media attention analysis. One related task is cross-document event coreference, which involves clustering mentions of the same event across a set of documents.
In comparison to our approach, it effectively builds an event database from the corpus. However, performance is typically poor and most methods cannot scale to large datasets \cite{eirew-etal-2021-wec}; indeed, many earlier methods are trained and evaluated on extremely small datasets, e.g.
% \ ECB+'s 
$502$ articles in the study of \citeauthor{cybulska-vossen-2014-using}~\cite{chen-etal-2023-cross,cattan-etal-2021-cross-document,barhom-etal-2019-revisiting,caciularu-etal-2021-cdlm-cross,choubey-huang-2017-event,held-etal-2021-focus,hsu-horwood-2022-contrastive,allaway-etal-2021-sequential}. While some newer datasets are larger, they are still orders of magnitude smaller than our article repository \cite{eirew-etal-2021-wec,eirew-etal-2022-cross}.
%For our article matching for events task, where location-dependent knowledge is critical (e.g., that \textit{Milford Sound} is in \textit{New Zealand}), gathering such training data will be very difficult. 
%saez2013social,
% Motivated by an earlier political science study, \citet{halterman-etal-2021-corpus} evaluate NLP methods to detect temporal trends in real-world police activity from a news corpus, but do not seek to link events across documents.

On the other hand, when social science researchers seek to analyze news bias, they typically use much simpler manual or heuristic methods to identify news articles discussing key events \cite{eisensee2007news,carron2013sagas,Hellmueller02012022,Dietrich01112020,baum2015reporting}, which is time consuming and hard to scale. Further, we find that previously used heuristic keyword matching approaches, when adapted to our task, are prone to significant error.

As the field of computational social science advances \cite{zuckerman2003global, Kwak2014UnderstandingNG,Hamborg2019}, new methods are being used to analyze large datasets, offering a more comprehensive view of international news coverage. 
%Techniques such as analyzing search engine query results have started to provide insights into the global priorities of news coverage \cite{}. 
Such studies promise to deepen our understanding of the mechanisms behind news selection and dissemination  globally. To this end, here we develop and evaluate a novel method for event identification, \textsc{FAME}, bridging the NLP and social science approaches.

\begin{figure*}
    \centering
    \includegraphics[width=1\textwidth]{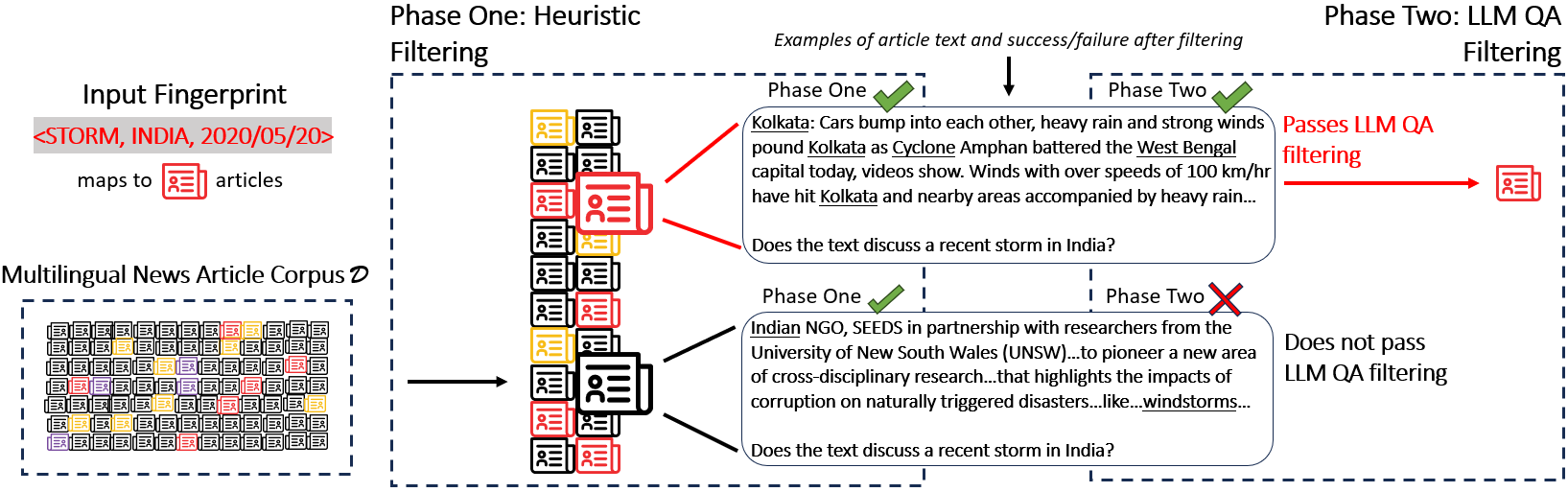}
    \caption{Examples of two articles from our dataset analyzed as candidates to link to specific event fingerprint $\langle$\textsc{Storm}, \textsc{India}, 2020/05/20$\rangle$. 
    For Phase One, matched keywords are underlined: from $K_{\color{red}\text{STORM}}$ = \{{storm, hailstorm, \underline{cyclone}, snowstorm, typhoon, \underline{windstorm}, rainstorm...}\} + suffixes; and 
      $K_{\color{red}\text{INDIA}}$ = \{{India, \underline{Indian}, \underline{West Bengal}, Uttar Pradesh, Tamil Nadu, Sikkim, Rajasthan, \underline{Kolkata}, Odisha...}\} + suffixes.
      LLM QA prompt: \textit{Does the text discuss a recent \textcolor{red}{storm} in \textcolor{red}{India}?} }

      \label{fig:method}
\end{figure*}

\textbf{Agenda setting and news about disasters and terrorism.} The media plays a crucial agenda-setting role in reporting disasters and terrorist attacks. News organizations act as gatekeepers, deciding which events gain public attention, thereby shaping the ``agenda" of public concerns. \cite{shoemaker2009gatekeeping, mccombs1972agenda, mccombs2021setting}.

Research highlights several factors influencing the newsworthiness of disasters and terrorist attacks. While death tolls are important predictors, proximity—cultural, geographic, and economic—also plays a key role in determining media attention \cite{wu2007brave, greer2003sex}. Cultural affinity, geographic closeness, and economic ties have all been shown to influence coverage \cite{adams_whose_1986, berlemann_distance, joye_news_2010}. While these findings are robust, most studies examine US media outlets, with limited research on non-American sources.

Terrorism reporting has unique dynamics, as terrorism inherently aims to attract media attention. Cultural factors significantly affect agenda, as seen in US media’s disproportionate focus on Muslim perpetrators \cite{Kearns19092019}. Comparative studies reveal differences in news selection criteria; for example, Chinese media are more selective in covering terrorist attacks compared to US outlets \cite{zhang_reality}.

\textbf{Most related work.} In this study, we focus on conceptual replication and broadening of \citet{eisensee2007news}'s seminal analysis of US\ television news attention toward global disasters. 
This study shows that a disaster in Africa requires significantly higher death tolls to gain equivalent coverage to one in Eastern Europe in US media.
We extend this research to a much larger and multilingual corpus, analyze the news coverage in the US and outside of it, and study both coverage of disasters \textit{and} terrorist attack events to illustrate our method's flexibility.
Finally, we provide the first evaluation of \citet{eisensee2007news}'s keyword-based article matching method, finding significant errors, motivating the  development of an alternative.

% \textbf{Most similar work.} We focus on conceptual replication and broadening of \citet{eisensee2007news}'s analysis of US\ television news attention toward global disasters, which carefully considers attention effects across a very large number of countries. 
%  We extend to a much larger and multilingual corpus, and also add terrorist attack events to illustrate the method's breadth.
% We provide the first evaluation of their keyword-based article matching method, finding significant errors, motivating the  development of an alternative.

\section{Linking News Articles to Event Records}
\label{s:linking}

In this section, we first introduce the fundamental task of identifying news discussing a particular event, using event metadata from databases of event records. Given the large datasets of event records and articles that we aim to link, a methodology that can solve this fundamental task should scale to hundreds of millions of news articles across different languages and thousands of events that naturally happen in our world.

\subsection{General Problem Statement}\label{sec:problemstatement}

Event information exists in an event database, $\mathcal{E}$, whose records contain metadata about each event---in particular, what it is, and where and when it happened. We formulate an event's metadata as a \textit{fingerprint}, $e \in \mathcal{E}$. Given event fingerprints, and a large corpus of time-stamped news articles, $\mathcal{D}$, the fundamental task of any event coverage analysis is to identify news articles that describe each of the events---that is, for every $e \in \mathcal{E}$, to retrieve the set of documents, $D_{e} \subset \mathcal{D}$, that discuss an event specified by fingerprint $e$,
for any $e \in \mathcal{E}$. Figure~\ref{fig:linking} shows an example of the input and desired output for this task.

We next describe the news document dataset ($\mathcal{D}$) and event databases ($\mathcal{E}$) that we focus on in this study.

\subsection{Large Dataset of News Articles}\label{sec:datasetnews}
As a large corpus of news articles, $\mathcal{D}$, we use data from Media Cloud, a platform that collects articles from news feeds of 1.1 million distinct media outlets since 2008~\cite{roberts2021media}. The platform maintains a list of news outlets organized by country\footnote{https://sources.mediacloud.org/\#/collections/country-and-state}, which has been carefully curated over years by international media scholars to cover important news outlets~\cite{roberts2021media}.
In this study, we exclude any URLs from popular social media platforms (such as twitter.com, facebook.com, reddit.com).
We collected metadata and full text of all, $\sim$42 million, news articles published from January 1, 2020 to June 30, 2020 in three languages~\cite{chen-etal-2022-semeval, xi_global_news_2024}: 
English (31M articles),
Spanish (8.2M), and
French (3.2M).
We chose these languages because they are among the top 6 languages with most speakers worldwide~\cite{eberhard2021top} and our team includes members proficient in these languages, which facilitated data exploration. 
% \bto{we could mention that they're well-represented within mediacloud.  by data volume, they're all within top-10 i thought...} 

\begin{table*} [ht]{\small
\centering
\begin{tabular}{cccccccc}
& & \multicolumn{3}{c}{Passing Phase 1}
& \multicolumn{3}{c}{Passing Phase 2}
\\
\cmidrule(lr){3-5}\cmidrule(lr){6-8} 
&&
& \multicolumn{2}{c}{over events}
&& \multicolumn{2}{c}{over events} \\
\cmidrule(lr){4-5}\cmidrule(lr){7-8} 
Lang. & Total & Total & Median & Max & Total &  Median & Max  \\
\midrule 
EN & 31 M  & 82268 & 89 &  6082  & 18658  &  13 &  2405  \\
ES & 8.2 M  & 61460& 63 &  2771 &   7506 & 1.5 &  1000  \\
FR & 3.2 M  & 8571 & 8 &  427 & 1277 &  0  & 167 \\
%EN & 31 M  & 89 &  6082  & 82268 & 13 &  2405 & 18658 \\
%ES & 8.2 M  & 63 &  2771 & 61460& 1.5 &  1000 & 7506 \\
%FR & 3.2 M  & 8 &  427 & 8571 & 0  & 167 & 1277\\
\hline
\end{tabular}
\caption{For each language's subcorpus: the total number of initial articles, and how many pass each phase of \textsc{FAME}'s pipeline; and the median and maximum number of articles matched per event.
}\label{tab:articlestats}}
\end{table*}

\subsection{Databases of Event Records}\label{sec:eventrecords}

We aim to identify news articles about critical events based on event records from databases, $\mathcal{E}$, of disasters (EM-DAT, USGS) and terrorist attacks (GTD). EM-DAT\footnote{https://www.emdat.be/} is a comprehensive database storing details about natural disasters, compiled from UN agencies, research institutes, and press agencies, among others. We focus on natural disasters that include hydrological, geophysical, and meteorological disasters. The United States Geological Survey\footnote{https://earthquake.usgs.gov/earthquakes/map/} (USGS) is a database storing details about earthquakes of magnitude 4.5 or higher throughout the world as part of the National Earthquake Hazards Reduction Program (NEHRP). %, which is a four-agency partnership established by the US Congress. 
% It receives data on earthquakes from observatories and geophysical institutions throughout the world and typically locates and reports earthquakes of magnitude 5 or higher within 30 minutes.
The Global Terrorism Database\footnote{http://apps.start.umd.edu/gtd/} (GTD) is a comprehensive and open-source database that contains information about a variety of terrorist incident events around the world from 1970 through 2020, where information is based on a variety of open media sources. 
% The National Consortium for the Study of Terrorism and Responses to Terrorism (START) makes the GTD available online. 
Since GTD includes many terrorist incidents, we focus only on the most salient events that result in more than 10 casualties or constitute one of the top $3$ terrorist attacks in terms of casualties for each unique country. 

After this selection from the event databases, we accumulate $188$ natural disaster and $282$ attack events that started and finished in the first six months of 2020 for our news coverage study.

\section{Method Identifying News on Critical Events}
\label{s:method}

We introduce and use a method that efficiently links multilingual news articles in $\mathcal{D}$ to records that are listed in an event database, $\mathcal{E}$.

\subsection{Input and Fingerprint Specification}\label{sec:inputfingerprint}

\textsc{FAME} takes as input an event fingerprint, $e\in \mathcal{E}$, which is a narrow subset of event record metadata, and a dataset of news articles $\mathcal{D}$. The fingerprint is a triplet $e=\langle c, l, t \rangle$  (e.g., $\langle\textsc{Storm},\textsc{India}, 2020/05/20\rangle$), where:
% \begin{itemize}
    % \item 
    $c\in \mathcal{C}$ encodes an event class (e.g., \textsc{Storm}),
    % \item 
    $l \in \mathcal{L}$ is a location of an event (e.g., \textsc{Chile}),
    % \item 
    and $t$ is the starting date of an event    (e.g., 2020/05/17). 
% \end{itemize}
%Figure~\ref{fig:method} shows a real fingerprint from EM-DAT, $\langle\textsc{Storm},\textsc{India}, 2020/05/20\rangle$. 
Here, $\mathcal{C}$ is a set of discrete event classes (e.g., \textsc{Attack}, \textsc{Storm}) and $\mathcal{L}$ is a set of discrete location classes (e.g., \textsc{India}, \textsc{Italy}). Such basic metadata is widely available across event record databases. More detailed metadata or types of event arguments, such as an \textsc{Attack}'s perpetrator group, is not in the fingerprint because they may be too class-specific, and can raise broader semantic issues; we leave incorporation of these for future work.

These are the same inputs as the ones used in the study of \citet{eisensee2007news}, which we seek to replicate and expand.
This input specification assumes that there are no duplicate event fingerprints, i.e., two events having the same fingerprint. This assumption holds true for all natural disasters and for $98\%$ of the terrorist attacks for the databases we work with.

\subsection{Method}\label{sec:method}

Our method for extracting multilingual articles about events follows the following steps. (1) \textit{Heuristic filtering} selects a set of candidate articles that might discuss the event of interest. We apply a rule-based approach to select these candidate articles and aim to preserve recall in this step.
(2) \textit{LLM filtering} selects a final subset of articles. Here, we apply LLM QA. We illustrate an example application of heuristic filtering and LLM filtering in Figure~\ref{fig:method}. 

\subsubsection{Phase One: Heuristic filtering}\label{sec:phaseone} For each event fingerprint, the heuristic filtering step aims to prune the original set of articles to a small subset which  includes all articles that discuss the event specified by the fingerprint, as in Figure~\ref{fig:method}. Given fingerprint $\langle c, l, t \rangle$, the filtered documents must satisfy three rules:
\begin{itemize}
\item[(1)] Be within seven days of the start date $t$ of the event, \item[(2)] Contain a keyword indicating the location of the event ($k_{l}\in K_{l}$ where $K_{l}$ is a set of keywords that may indicate the occurrence of an event in location $l\in\mathcal{L}$ (e.g., Kolkata, Odisha... for \textsc{India})), 
\item[(3)] Contain a keyword indicating the class of the event ($k_c\in K_c$ where $K_c$ is a set of keywords that may indicate the occurrence of an event of class $c\in \mathcal{C}$ (e.g., blizzard, typhoon... for \textsc{storm}). 
\end{itemize}

\noindent Figure~\ref{fig:method} shows examples of document text that passes Phase One heuristic filtering for the fingerprint, $\langle$\textsc{STORM}, \textsc{INDIA}, 2020/05/20$\rangle$.

To increase the chance that all articles discussing an event described by fingerprint $\langle c, l, t \rangle$ pass this Phase One step, $K_c$ and $K_{l}$ must be comprehensive keyword sets, containing as many keywords as possible that could indicate the occurrence of the event. To construct $K_c$ and $K_{l}$, we used glossaries and lexical databases such as WordNet \cite{miller1995wordnet}, which stores synonyms and hyponyms---words that store more specific semantic fields. Then, following the approach of \citet{cai2024montecarlolanguagemodel}, we expanded the sets using an LLM by setting its temperature hyperparameter---which controls the extent of output randomness---to 0.5 and repeatedly prompting for candidate synonyms and hyponyms of $c$ and $l$. We ultimately added any words that appeared in at least $50\%$ of the LLM outputs to the keyword sets.

\subsubsection{Phase Two: LLM filtering} Since Phase One extracts documents that may potentially discuss an event specified by a fingerprint, aiming for perfect recall but not high precision, a second filtering step uses LLM QA to link event fingerprints with articles, with high precision and recall. 
For each candidate article that passes initial filtering, \textsc{FAME} sends to GPT-3.5-turbo the following \textit{LLM prompt}:

% \noindent
% \textbf{LLM Prompt:}

\vspace{1em}
\fbox{\begin{minipage}{21em}\texttt{[article title + first 3 sentences]}\\

\noindent\texttt{Does the text discuss a recent [event class $c$] in [location $l$]?}
\end{minipage}}
\vspace{2em}

\noindent Answers from the LLM that include a `\textit{Yes}' response indicate that an article discusses an event; otherwise, the article is dropped. We finalized the question posed to the LLM after evaluating four versions of questions that varied based on specificity (Appendix~\ref{app:ablation}). The LLM prompt contains only the article title and first $3$ sentences of article text because we found that out of a sample of $150$ articles across various event classes, if an article positively mentioned an event, it mentioned the event in either the title or first $3$ sentences.

We use GPT-3.5-turbo for the LLM QA step, even though other large language models, including later versions of this one, are available because it performs well and is cheap. In our experiments, open source models did not perform as well (results are in Appendix~\ref{app:ablation}), but we hope to explore this possibility in our future research.

\subsection{Handling of Multiple Languages}\label{sec:applytolangs}

\textsc{FAME} can extract articles that discuss an event specified by a fingerprint for any language. For the Phase One heuristic filtering step, the keyword sets $K_c$ and $K_l$ consist of words that belong to the same language as those in the text documents. We populate the keyword sets in all languages (English, Spanish, and French) using the same procedure (\S\ref{sec:mapping}). 

While the Phase One heuristic filtering step requires keyword engineering for each language, the Phase Two LLM QA step applies cross-lingual QA, with no customization for each language. Specifically, the LLM prompt uses the article title and first 3 sentences of article text, word-for-word, and poses a question in English irrespective of the language of the article text. This cross-lingual approach does not face prompt and wording sensitivity issues for questions in languages that are not English, and follows a vast literature supporting cross-lingual QA \cite{liu-etal-2019-xqa,lewis-etal-2020-mlqa,muller-etal-2022-cross,zhou-etal-2021-improving}. We additionally find that it performs well for linking events with Spanish- and French-language articles in~\S\ref{s:performance}.

\section{Identifying News about Disasters and Terrorist Attacks}
\label{s:eval}
% \section{Identifying news about critical events}

Here, we first describe how \textsc{FAME} is adjusted to the event databases by choosing relevant keywords sets $K_l$ and $K_c$.
Then, we apply \textsc{FAME} to identify news about natural disasters and terrorist attacks. We label 6,468 
% $2514+ 2361+ 1593$
event-news pairs passing Phase One as either correct or incorrect matches.
Using these labeled samples as a ground truth, we evaluate \textsc{FAME}'s performance across eight event classes.
% (e.g., \textsc{storm}, \textsc{attack}, \textsc{flood}).

\subsection{Adjusting \textsc{FAME} to the Event Databases}\label{sec:mapping}

For our evaluation and social science study, the event fingerprints $\langle c,l,t\rangle$ are from EM-DAT, USGS, and GTD databases, introduced in \S\ref{s:linking}.
The set of discrete event classes $\mathcal{C}$ includes natural disaster and attack classes: \textsc{earthquake}, \textsc{flood}, \textsc{avalanche}, \textsc{storm}, \textsc{landslide}, \textsc{volcano}, \textsc{wildfire}, and \textsc{attack}, and the set $\mathcal{L}$ of discrete location classes corresponds to countries, all derived from EM-DAT, USGS, and GTD.

Keyword sets $K_l$ and $K_c$ consist of location and event class keywords respectively, and aim to be comprehensive. Therefore, $K_l$, which is the keyword set corresponding to country $l\in\mathcal{L}$, consists of the country name and demonyms from the Open Event Data Alliance's \texttt{CountryInfo.txt},\footnote{https://github.com/openeventdata/CountryInfo} as used in \cite{OConnor2013IR}
%\footnote{\scriptsize \url{https://github.com/brendano/OConnor_IREvents_ACL2013}}, 
province names as indicated by `admin1' in the Geonames database\footnote{https://www.geonames.org/}, and any city in the top 5000 most populated cities in the world as per the GeoNames database. These databases are all in English; we used Google Translate\footnote{https://translate.google.com/} to convert database entries to Spanish and French languages, and verified a sample of $200$ of the translations with fluent speakers. $K_c$, which is the keyword set corresponding to event class $c$, consists of synonyms or hyponyms  which are from thesauruses\footnote{https://www.merriam-webster.com/thesaurus/, https://dictionary.cambridge.org/dictionary/english-french/, https://dictionary.cambridge.org/dictionary/english-spanish/}, are sampled from GPT-3.5-turbo, or are derived from the WordNet lexical database. These are stemmed and appended to different affixes (e.g., -s, -ing, etc.). Since WordNet is an English lexical database, we used Google Translate and verified the Spanish and French keyword sets with fluent speakers.

\subsection{How Many Articles Pass \textsc{FAME} Filters?}

Next, we apply \textsc{FAME} to the three databases. 
In Table~\ref{tab:articlestats}, we present the number of articles passing each Phase of FAME, including their statistics per event.
% information on the resulting $|D_e|$ distribution across all $470$ events.

\noindent Phase One drastically filters down articles by more than
two orders of magnitude,
%> c(31e6, 8.2e6, 3.2e6) / c(82268, 61460, 8571)
%[1] 376.8172 133.4201 373.3520
highlighting its importance for computational efficiency---the initial corpus is far too large to run through an LLM.
Then in Phase Two, LLM QA again removes a large number of articles (more than three quarters), 
%> 1-c(18657, 7506, 1277) / c(82268, 61460, 8571)
%[1] 0.7732168 0.8778718 0.8510092
motivating the next section's manual evaluation to assess if it is removing false positives as intended.
Consistent with \citet{zuckerman2019whose}, we also find that most events have very low news coverage---in fact, for all phases and each language's subcorpus, there are events with zero linked articles (i.e., our method finds no news coverage),
while a small number of events have significantly higher news coverage.

\subsection{Ground Truth Event-news Pairs}

We label pairs of events and news identified by \textsc{FAME} in English, Spanish and French over all $8$ event classes.  
If more than $3$ event fingerprints in $\mathcal{E}$ have the same event class, we randomly selected pairs corresponding to $3$ events from that class for labeling %event fingerprint triples of the event class from EM-DAT, USGS, and GTD 
($4$~for the \textsc{Attack} event class, of which there are significantly more events), where each event fingerprint  corresponds to at least 30 articles passing Phase One, and at least 5 articles passing Phase Two. 
If fewer than $3$ event fingerprints in $\mathcal{E}$ have the same event class, we labeled pairs of events and news corresponding to all events of that class. 
%This allows us to avoid evaluation on very small sets of articles. For event classes that correspond to $\leq 3$ event fingerprint triples in $\mathcal{E}$, we selected all of the triples corresponding to the class.  
We compute precision, recall and F1 scores for each language and each event fingerprint that we labeled news for.

We construct ground truth labels of articles corresponding to event fingerprint records $\langle c, l, t \rangle$ using annotators hired from Upwork who are fluent in the target languages.\footnote{Each annotator has between $98$ and $100$ percent satisfaction over their previous jobs. Compensation was roughly \$25 per hour. The UMass Human Research Protection Office determined this annotation study is not human subjects research.}  For each language, two annotators labeled all or the first 150 Media Cloud articles that passed the Phase One heuristic filtering step, which aimed to have perfect recall, for each event record. An article received a positive label if any part of its title or first three sentences of text discusses the input event record and a negative label otherwise, indicating no or peripheral discussion of the event. Annotators for English-, Spanish-, and French-language text agreed $94.5\%$, $95.6\%$, and $95.8\%$ of the time respectively (Cohen's kappa of $0.88$, $0.91$, and $0.91$ respectively). Most disagreements (approximately $90\%$) stemmed from inadvertent errors, such as overlooking relevant text, inconsistent application of the annotation criteria, and skipping annotation of certain articles entirely. Disagreements were resolved by the authors through a systematic review of the annotation guidelines. To ensure consistency, all resolved labels were carefully aligned with the predefined annotation criteria.

Given that the ground truth data consists of articles that pass Phase One, we seek as complete a set of articles as possible, using a very large set of keywords containing synonyms and hyponyms. We test whether the keyword sets are expansive enough by performing an ablation that considers even more distant synonyms in the keyword sets (details in Appendix~\ref{app:ablation}). Because we find that \textsc{FAME} links the same articles, we believe our original keyword sets are as large, relevant and varied as they need to be, though we cannot guarantee perfect recall in retrieval of relevant articles from Media Cloud.

\subsection{Performance Comparison}
\label{s:performance}

\begin{table}{\footnotesize
\begin{center}
\begin{tabular}{ccccc}\toprule
\textbf{Lang.}& \textbf{Method} & \textbf{Prec.} & \textbf{Recall} & \textbf{F1} \\ \midrule%\cmidrule(lr)[4-4]\cmidrule(lr)[5-5]
EN & KW: Title-only & 88.1 & 48.4 & 57.6 \\
EN & KW: Title+Body & 84.1 & 78.0 & 79.3 \\
\midrule
EN & \textsc{FAME} & 93.2 & 95.4 & 94.0 \\
FR & \textsc{FAME} & 96.9 & 94.1 & 95.2 \\
ES & \textsc{FAME} & 98.1 & 96.0 & 96.8 \\
\bottomrule
\end{tabular}\end{center}\caption{Averaged precision, recall, and F1 scores over expert-annotated articles matched to events for \textsc{FAME} and two previously used keyword classifier baselines ("KW") adapted to our task on English data (title-only, and on first three sentences of article text).
% that do not have zero prevalence, 
Results for English (\textsc{FAME} and the baselines), Spanish (\textsc{FAME}), and French (\textsc{FAME}). }
\label{tab:performance}}

\end{table}

% BELOW: old fro brendan approx Jan10
% once averaged results table is here i would rewrite this section as:
% ..discuss averaged results..
% but, while overall broad results evaluations are typically used in NLP, they may mask significant variation across events, which are ultimately our units of analysis.  therefore we show models' precision performance broken out over each sample [?] of events.  note ....  possible future work ....
%
% The point is the per-event breakdown isn't as fundamental - it's an add-on so should come second.  spanish/french results are NOT an add-on, they're fundamental to our main argument and the case study section!!

The averaged results of our evaluation in Table~\ref{tab:performance}  highlight the high performance of our method for linking \mbox{English-,} Spanish-, and French-language articles to events.\footnote{Averaged events must have at least one article discussing it to properly calculate precision and recall.} Since many social science studies use keyword matching to identify event coverage \cite{Hellmueller02012022}, we adapt the method in \citet{eisensee2007news}, which has the most similar task definition as \textsc{FAME}, as a baseline, using the same location and event class keywords that they use. Since \citet{eisensee2007news} does not analyze or have keywords for \textsc{Attack} events, we construct the keyword set based on the attack hierarchy and definitions in the GTD codebook. We do not use existing NLP methods as baselines since they either rely on large amounts of training data to perform well or face other inconsistencies with our task as discussed in \S\ref{s:relatedwork}.

Specifically, for English-language articles, our baselines consist of keyword matching on Media Cloud article titles, and on article titles along with the first three sentences of article text. In both cases, the approach produces significant errors and performs more poorly than \textsc{FAME}.

\begin{table*}[ht]{\small
\centering
\begin{tabular}{l|ccc|rrr}
\toprule
\textbf{Event} & \textbf{Class} ($c$) & \textbf{Location} ($l$) & \textbf{Time} ($t$) & \textbf{News} & \textbf{Death} & \\
\midrule
Cyclone Amphan & Storm & India & May 20 & 2723 & 90 \\
Taal Volcano eruption &Volcano & Philippines & Jan 12 & 2617 & 1 \\
Magnitude 6.4 quake & Earthq. & Puerto Rico & Jan 7 & 1690 & 2 \\
Cyclone Amphan & Storm & Bangladesh & May 20 & 1437 & 26 \\
Storm Ciara & Storm & UK & Feb 8 & 1314 & - \\
Severe thunderstorms & Storm & USA & Mar 2 & 1268 & - \\
2020 Elazığ quake & Earthq. & Turkey & Jan 24 & 1100 & 41 \\
Storm Gloria & Storm & Spain & Jan 19 & 1046 & 4 \\
Severe Storm & Storm & USA & Apr 10 & 857 & -\\
Storm Amanda & Storm & El Salvador & May 31 & 835 & - \\
\bottomrule
\end{tabular}
\caption{Top 10 disasters with the most news in the first five months of 2020. Event names were derived from the headlines of news or titles of respective Wikipedia pages.}
\label{tab:top_disasters_2020}}
\end{table*}

While aggregate results as in Table~\ref{tab:performance} are typically used in NLP, they may mask significant variation across events, which are ultimately our units of analysis. Therefore, we show models' precision and recall performance broken out over each event in Figure~\ref{fig:performance} (the results for Spanish in French are in Appendix~\ref{app:per-event-eval-es-fr}).  
We note that while the averaged KW:Title+Body result, of $79.3$ F1, may sound reasonably accurate, this masks massive variation---some events see precision low as $21.4$ (and recall, $26.9$)!  This could have negative consequences when trying to make qualitative or social scientific interpretations from LLM/NLP output, highlighting the need for more NLP evaluation---and perhaps new modeling approaches---that better target the finer levels of substantively relevant granularity that are important for computational social science.

\subsection{Ranking of Events with Most News Coverage}
To qualitatively evaluate the results of \textsc{FAME}, we examine the events that experienced the greatest news coverage in the first half of 2020. Table~\ref{tab:top_disasters_2020} presents the 10 disasters that garnered the highest number of news reports as retrieved by \textsc{FAME}. The majority of these events are documented on Wikipedia. Notably, our method successfully identifies events associated with the same disaster occurring in different locations, such as Cyclone Amphan which affected India and Bangladesh. For space considerations, the top 10 terrorist attacks are provided in Appendix~\ref{app:reporting-countries-examples-of-attacks}.

\section{Investigating Bias in News Coverage}
\label{s:regression}

Event identification allows us to examine how news coverage differs across countries. Prior work found that a disaster in Africa requires a death toll that is 40 times higher than one in Eastern Europe to receive equivalent news coverage in US media \cite{eisensee2007news}. However, it is unclear whether such differences in media coverage are due to geographic distance between the countries where the event occurred and where it was reported, or duo to other factors, such as cultural, societal, or economic proximity.

Using articles extracted from Media Cloud from between January and May of 2020 using \textsc{FAME}, we perform a large-scale investigation of patterns in news coverage of natural disasters and terrorist attacks. First, we compute the average number of news stories per event in a country reported in another country.\footnote{We consider only the events with unique fingerprints.} Then, to identify patterns in media coverage of such events, we regress this average against a set of 47 country characteristics proposed as determining factors by media scholars. We were able to gather these characteristics for 128 countries, which form the base for our regression analysis, since samples in our regression models correspond to country pairs.

%Then, we leverage country characteristics that media scholarship hypothesized as factors determining news coverage as predictors in regression models of \textit{the average number of news} reported per disaster or terrorist attack. 

We describe details of our regression models in the next subsection (\S\ref{sec:reg_details}).
Then, we analyze the US media coverage of critical events (\S\ref{sec:reg_us}). Finally, we expand the scope to investigate global media coverage of disasters and terrorist attacks (\S\ref{sec:reg_us}), focusing on reports from countries where either English, Spanish, or French is an official language.

\subsection{Details of Regression Models}
\label{sec:reg_details}

\paragraph{Sample selection}
Overall, our set of events happen in 128 unique countries. If a country did not report on an event, then the dependent variable for that country pair is zero. To limit the number of zeroes, we restrict reporting countries either to the US (\S\ref{sec:reg_us}), or to the countries that cover critical events happening in at least ten countries (\S\ref{sec:reg_global}). There are 10 such reporting countries in English, 4 in Spanish, and 2 in French (full list in Appendix~\ref{app:reporting-countries-examples-of-attacks}). 

% We conducted a literature review 
\paragraph{Choice of candidate factors}
To identify factors that might influence news coverage of disasters, we conducted a literature review (\S\ref{s:relatedwork}). We classified the factors into five categories:
event characteristics, economic, societal, geographic and cultural \cite{lee2007international, segev2016group}. We consider key characteristics of the specific event, such us the number of deaths, and the characteristics of the country where that event happened: its continent, GDP, Gini index of income inequality, and Democracy index\cite{economist_democracy_index}. Country indices are sourced from the World Bank and The Organization for Economic Co-operation and Development (OECD). Most of the factors we consider are relational, that is they describe the relationship between the country where the critical event happened and the country where it might be covered, e.g., whether these countries are neighbors, share an official language, their levels of trade and direct investment in each other. All 47 considered factors are listed in Appendix~\ref{app:country-factors-regression} with their detailed explanations, definitions, and value ranges.

\begin{figure}
    \centering
    \includegraphics[scale=0.34]{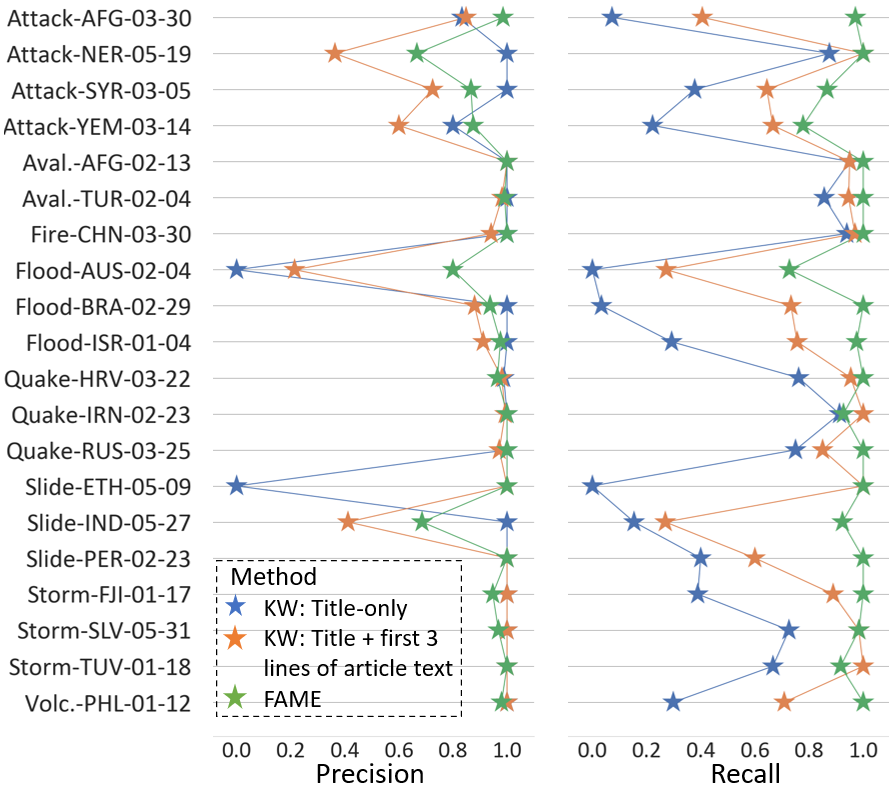}
    \caption{Precision and recall of \textsc{FAME} and the two baselines (``KW") for random events of each event class.
    % on extracting articles about randomly selected disaster events of different classes.
    }
    \label{fig:performance}
\end{figure}

\paragraph{Feature and model selection}
Next, we evaluate which of the 47 candidate factors may be related to the news coverage of disaster and terrorist attack events. Given the possibility of collinearity and irrelevance among these factors, we employed a feature selection algorithm to discern the most impactful predictors. Specifically, we identify the features that are the most significant through forward selection to greedily optimize the Akaike Information Criterion (AIC), adding factors one by one. The AIC method assists in identifying the optimal set of predictors that most effectively explain the variations in the dependent variables, ensuring a robust and effective model. We also experimented with Variance Inflation Factor (VIF) feature selection, which yielded similar results.
% In addition to factors identified by the literature, we took into account two additional factors: the Internet user rate, to account for the relevance of online news in today's media landscape.

% To avoid co-linearity, we applied feature and model selection techniques. Namely, we employ two well-known criteria---Variance Inflation Factor (VIF) and Akaike information criterion (AIC)---and report results for the best models according to each criterion. We report models constructed with either of the techniques to showcase robustness of our results and methodology. VIF is a popular feature selection technique, whereas AIC is a model selection method that identifies the predictors that best explain the dependent variables. For all regressions we apply the same model and feature selection.

\paragraph{Factor preprocessing}
To ensure comparability across factors, each factor was normalized using a min/max scaler, which rescales every factor to a numeric value between 0 and 1. Additionally, logarithmic scaling was applied to the death count for smoothing the skewness of the data. Death count information is missing for some events: in such cases we imputed it with the average death count per event, following our manual examination of a subset of 20 disasters without death counts.\footnote{For about a half of them we inferred non-zero death counts based on information available in the Internet, but we could find the exact death count only for three of them.}. 
% For the analysis, we restricted our observations to reporting countries (10 in English, 4 in Spanish, and 2 in French; full list in Appendix~C) that cover the news events in at least ten countries, thereby mitigating variance within the samples. 
% We considered disaster and terrorist attack events with explicitly recorded death counts and imputed missing values using the average death numbers. This approach ensures the robustness and consistency of our regression analyses.

% Surprising result -- investment negative correlated to News coverage. Reasons can be complicated: (1) Countries that receive more coverage might attract more investment due to increased visibility and perceived stability or opportunities; (2) national media outlets pay less attention to the countries got investment for obtaining more interest (Countries might under-report negative events like disasters in countries where they have significant investments to avoid triggering economic instability or panic that could affect their investments). 

\subsection{US News Coverage}
\label{sec:reg_us}
We regress the average number of news articles published in the US about critical events happening in another country against country-level factors. Table~\ref{tab:us_disaster_conflict_merged} lists the coefficients of the factors identified by feature selection as predictive of the outcome variable. 

\begin{table}[htb]
\begin{center}
\resizebox{0.85\columnwidth}{!}{
\begin{tabular}{llr@{}lr@{}l}
\toprule
\textbf{Type} & \textbf{Variable} & \textbf{Disaster} & & \textbf{Attack} &\\
\toprule
Eve. & {\color{blue}\#Deaths}                 & \textbf{3.401}&$^{***}$ & \textbf{3.221}&$^{***}$ \\
\midrule
Geo. & {\color{Green} Neighbor}               & -2.234 & & -3.479 & \\
     & {\color{Green} Asia}                   & 0.382 & & 0.728 & \\
     & {\color{Green} Europe}                 & 0.472 & & 0.415 & \\
     & {\color{Green} Africa}                 & 0.335 & & 0.204 & \\
     & {\color{Green} Oceania}                & \textbf{1.615}&$^{***}$ & 0.310 & \\
     & {\color{Green} North America}          & \textbf{0.420}&$^{*}$ & 0.099 & \\
     & {\color{Green} South America}          & 0.207 & & 0.158 & \\
\midrule
Eco. & {\color{orange} GDP }                  & 0.315&$^{**}$ & \textbf{0.379}&$^{**}$ \\
     & {\color{orange} Trade}                 & 2.050 & & 4.427 & \\
     & {\color{orange} Investment}            & -0.254 & & 0.350 & \\
\midrule
Pol. & {\color{teal} Democracy in.}           & 0.060 & & -0.011 & \\
     & {\color{teal} Diplomatic relation}     & 0.080 & & -0.108 & \\
\midrule
Soc. & {\color{brown} Gini in.}               & -0.070 & & 0.032 & \\
\midrule
Cul. & {\color{purple} Same language}         & 0.042 & & -0.024 & \\
\bottomrule
     & const                                  & -0.442 & & -0.095 & \\
     & \#Observations                         & 128 & & 128 & \\
     & Adjust $R^2$                           & 0.603 & & 0.440 & \\
\bottomrule
\end{tabular}
}
\end{center}
\caption{Regression results of average US news coverage per disaster or terrorist attack happening in a foreign country. The top 3 significant coefficients with the largest absolute value are marked in bold font.
We indicate statistical significance at levels $p<0.001$ (***), $p<0.01$ (**), and $p<0.05$ (*).
}
\label{tab:us_disaster_conflict_merged}
\end{table}

First, the resulting models for disasters and terrorist attacks achieve high $R^2$ values, 0.603 and 0.440, meaning that these models explain 60\% and 44\% of the variance in the average number of news coverage per disaster, respectively.
We find that the factor that impacts the news coverage of disasters the most is the number of deaths (left column coefficients in Table~\ref{tab:us_disaster_conflict_merged}). This factor is also the most predictive for the news coverage of terrorist attacks (right column coefficients in Table~\ref{tab:us_disaster_conflict_merged}).  This common-sense relationship between mortality and coverage of disasters was described in prior works~\cite{eisensee2007news} and, thus, suggests \textsc{FAME} is retrieving relevant data.

Second, \citeauthor{eisensee2007news} (ES) find that geography matters significantly in the coverage of disasters in US news in the late 20th century. %~\cite{eisensee2007news}. 
We do not observe a significant bias in news coverage of disasters in Europe or Africa (left column in Table~\ref{tab:us_disaster_conflict_merged}), but we see that disasters in high GDP countries received significantly more attention. This is consistent with ES's provocative finding that disasters in Eastern Europe receive proportionally more coverage than disasters in Africa, since national GDP and GDP per capita are generally higher in Eastern European countries than in African countries. In our data,  a disaster in Nigeria requires 10 times more deaths than in Russia to receive the same level of attention, while a terrorist attack requires 21 times more deaths. While this similarity to ES is encouraging, there's a danger in comparing news from 2020 in our set to theirs, from 1968 to 2002. For most of ES's data, the Cold War meant that Eastern Europe was of particular interest to US audiences. In our data, we note a high degree of attention to disasters in North America and Oceania: wealthy nations (the US, Canada, Australia, New Zealand)  are well-represented on these continents.

%That said, the result of Eisensee is for news coverage of disasters between 1968 and 2002, which is a very different time period than the period of our study, that is the first half of 2020. In particular, Eastern Europe during the Cold War and after the fall of Soviet Union may have received more attention from the US media than it did in 2020. At the level of continents, during the first half of 2020 we observe significantly more US news coverage of disasters happening in North America and Oceania, which likely correspond to the presence of the US, Canada, and Australia in these regions.

\begin{table}[htb]
\begin{center}
\resizebox{0.95\columnwidth}{!}{
\begin{tabular}{llr@{}lr@{}lr@{}l}
\toprule
\textbf{Type} & \textbf{Variable} & \multicolumn{2}{c}{\textbf{English}} & \multicolumn{2}{c}{\textbf{Spanish}} & \multicolumn{2}{c}{\textbf{French}} \\
\toprule
Eve. & {\color{blue} \#Deaths}                 & \textbf{2.961}&$^{***}$ & \textbf{3.485}&$^{***}$ & \textbf{2.343}&$^{***}$ \\
\midrule
Geo. & {\color{Green} Continent sim}           & 0.004&& 0.128&& \textbf{0.106}&$^{*}$ \\
     & {\color{Green} Neighbor}                & 0.271&$^{**}$ & 0.171&& -0.084& \\
     & {\color{Green} Asia}                    & 0.157&$^{*}$ & 0.222&& 0.048& \\
     & {\color{Green} Europe}                  & 0.213&$^{**}$ & 0.302&$^{**}$ & -0.043& \\
     & {\color{Green} Africa}                  & 0.074&& 0.216& & -0.017& \\
     & {\color{Green} Oceania}                 & \textbf{0.830}&$^{***}$ & \textbf{0.574}&$^{***}$ & \textbf{0.303}&$^{**}$ \\
     & {\color{Green} North America}           & 0.141&& 0.282&$^{*}$ & -0.041& \\
     & {\color{Green} South America}           & 0.073&& -0.040&& 0.001& \\
\midrule
Eco. & {\color{orange} GDP (high)}             & && && -0.012& \\
     & {\color{orange} GDP (high-high)}        & \textbf{0.351}&$^{***}$ & \textbf{0.556}&$^{***}$ & -0.012& \\
     & {\color{orange} GDP (high-low)}         & 0.057&& -0.070&& & \\
     & {\color{orange} Investment}            & -0.300&& -0.622&& -0.023& \\
     & {\color{orange} Trade}                 & 0.499&& -0.643&& -0.038& \\
\midrule
Pol. & {\color{teal} Democracy in.}            & 0.044&& 0.051&& 0.011& \\
     & {\color{teal} Democracy in. (h-l)}      & 0.060&& && & \\
     & {\color{teal} Diplomatic relation}      & 0.031&& -0.043&& -0.009& \\
\midrule
Soc. & {\color{brown} Gini in.}                & -0.059&& -0.076&& -0.019& \\
     & {\color{brown} Gini in. (h-h)}          & 0.004&& && & \\
\midrule
Cul. & {\color{purple} Same language}          & 0.025&& 0.344&$^{***}$ & -0.006& \\
\bottomrule
     & const                                   & -0.133&& -0.197&& 0.017& \\
     & \#Observations                          & 1281&& 511&& 256& \\
     & Adjust $R^2$                            & 0.623&& 0.639&& 0.740& \\
\bottomrule
\end{tabular}
}
\end{center}
\caption{Regression results of average news coverage between countries per disaster in different languages. Missing regression coefficient indicates that the respective factor is irrelevant according to feature selection.}
\label{tab:merged_disaster_regression_impute_aver}
\end{table}

Third, terrorist attacks achieve much more coverage if there is a large volume of trade between the reporting country and the country affected. This factor was by far the most predictive of news coverage of terrorist attack events in the US, even more so than the number of deaths. This finding echoes findings in existing studies of news coverage \cite{segev2016group, wu2000systemic, xi_global_news_2024}.

\subsection{News Coverage Across Countries and Languages}
\label{sec:reg_global}

We also performed a regression analysis at a global scale, examining the average number of news articles about an event in a country against various country-level factors. Table~\ref{tab:merged_disaster_regression_impute_aver} shows the coefficients of the factors identified as predictive for news coverage of disaster events, categorized by language. Similar to the analysis of US news coverage, the global analysis finds that the number of deaths, GDP, and event location in Oceania are among the most significant factors influencing global media attention. A higher death toll corresponds to greater news coverage, and Oceania's high coefficient across languages likely reflects the prominence of high-GDP countries such as Australia and New Zealand, which play significant roles in the geopolitical landscape. This pattern aligns with studies on systemic biases in global media coverage, suggesting that wealthier nations receive more attention due to their economic influence, media infrastructure, and geopolitical importance~\cite{wu2000systemic, segev2016group, xi_global_news_2024}. Unlike the US-based study by Eisensee, which identifies a bias against Africa in disaster coverage, the global analysis indicates more balanced attention toward disasters in Africa, Europe, and Asia.

\begin{table}
\begin{center}
\resizebox{0.95\columnwidth}{!}{
\begin{tabular}{llr@{}lr@{}lr@{}l}
\toprule
\textbf{Type} & \textbf{Variable} & \multicolumn{2}{c}{\textbf{English}} & \multicolumn{2}{c}{\textbf{Spanish}} & \multicolumn{2}{c}{\textbf{French}} \\
\toprule
Eve. & {\color{blue} \#Deaths}                 & \textbf{2.695}&$^{***}$ & \textbf{2.646}&$^{***}$ & \textbf{2.423}&$^{***}$ \\
\midrule
Geo. & {\color{Green} Continent sim}           & -0.015&& -0.099&& -0.016& \\
     & {\color{Green} Neighbor}                & 0.298&$^{**}$ & 0.024&& -0.056& \\
     & {\color{Green} Asia}                    & 0.311&$^{***}$ & 0.345&$^{*}$ & 0.280& \\
     & {\color{Green} Europe}                  & 0.251&$^{**}$ & 0.390&$^{**}$ & 0.261& \\
     & {\color{Green} Africa}                  & 0.076&& 0.261&& 0.407&$^{*}$ \\
     & {\color{Green} Oceania}                 & 0.171&& 0.618&$^{**}$ & \textbf{0.656}&$^{**}$ \\
     & {\color{Green} North America}           & 0.192&& 0.330&& 0.287& \\
     & {\color{Green} South America}           & 0.106&& \textbf{1.106}&$^{***}$ & 0.326& \\
\midrule
Eco. & {\color{orange} GDP (high)}             & 0.324&$^{***}$ & 1.012&$^{***}$ & 0.587&$^{***}$ \\
     & {\color{orange} GDP (high-high)}        & \textbf{0.561}&$^{***}$ & 0.490&$^{***}$ & 0.567&$^{***}$ \\
     & {\color{orange} GDP (high-low)}         & && -0.489&$^{***}$ & & \\
     & {\color{orange} Investment}            & 0.127&& 0.916&$^{*}$ & 0.560& \\
     & {\color{orange} Trade}                 & \textbf{1.967}&$^{***}$ & \textbf{3.858}&$^{***}$ & \textbf{1.798}&$^{**}$ \\
\midrule
Pol. & {\color{teal} Democracy in. (h-h)}      & -0.073&& -0.025&& -0.000& \\
     & {\color{teal} Democracy in. (h-l)}      & 0.155&$^{*}$ & -0.489&& & \\
     & {\color{teal} Diplomatic relation}      & 0.217&$^{***}$ & 0.318&$^{***}$ & 0.049& \\
\midrule
Soc. & {\color{brown} Gini in. (h-h)}          & 0.031&& 0.087&& -0.049& \\
     & {\color{brown} Gini in. (h-l)}          & 0.002&& && -0.049& \\
\midrule
Cul. & {\color{purple} Same language}          & 0.108&$^{**}$ & -0.205&& -0.247&$^{***}$ \\
\bottomrule
     & const                                   & -0.206&$^{*}$ & -0.400&$^{**}$ & -0.197& \\
     & \#Observations                          & 1281&& 511&& 256& \\
     & Adjust $R^2$                            & 0.484&& 0.518&& 0.499& \\
\bottomrule
\end{tabular}
}
\end{center}
\caption{Regression results of average news coverage between countries per terrorist attack in different languages.}
\label{tab:conflict_regression_merged}
\end{table}

Table~\ref{tab:conflict_regression_merged} presents the coefficients for terrorist attack events, demonstrating that the number of deaths remains the most predictive factor for news coverage across languages. Economic factors, however, exhibit a nuanced role in the context of terrorist attacks. While GDP remains significant, trade volume emerges as an equally predictive factor, underscoring the economic interdependence between nations and its influence on media agendas. Diplomatic relations also appear as a significant factor for terrorist attack events but not for disasters, suggesting that geopolitical alliances and strategic partnerships amplify media attention to certain attacks. South America exhibits a notable coefficient exclusively in Spanish-language media, while Africa appears prominently in French-language media. This pattern may reflect the heightened attention of the Spanish-speaking and French-speaking worlds to the respective continents' political instability and humanitarian crises.

\begin{figure*}[t!]
\centering
\includegraphics[width=0.98\columnwidth]{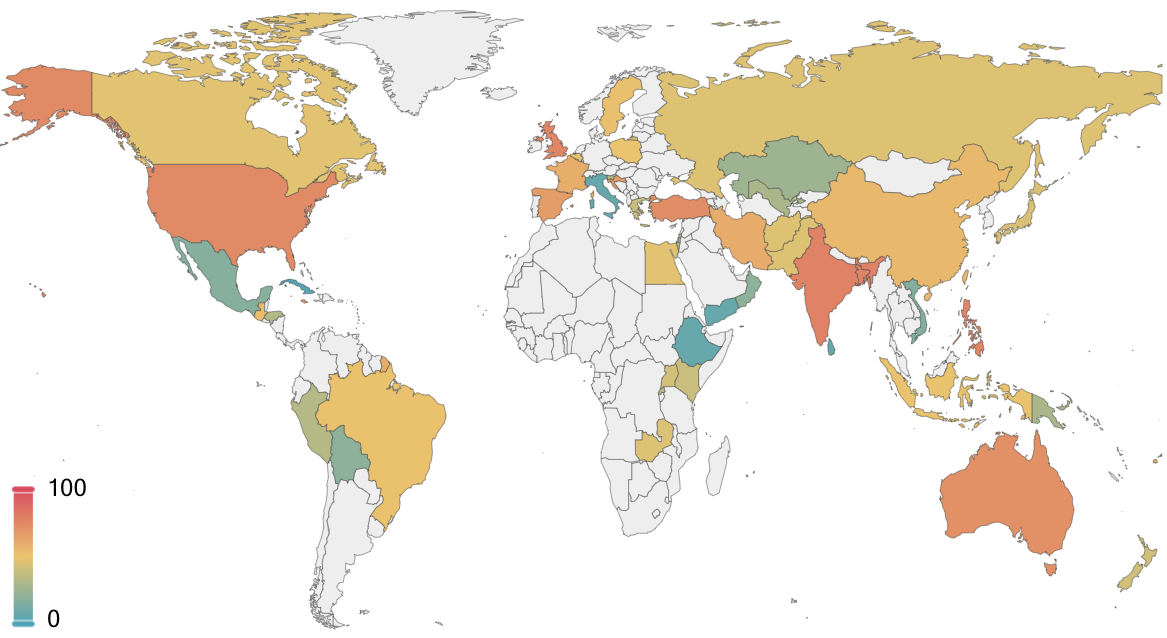}
\includegraphics[width=0.98\columnwidth]{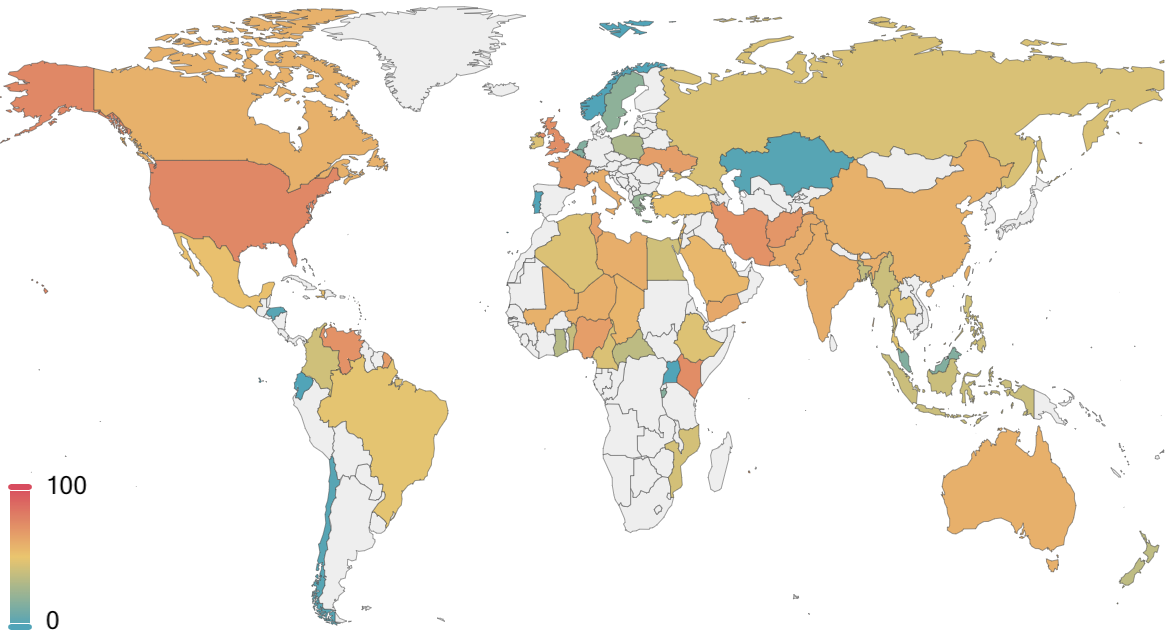}
\caption{The average number of news in English per disaster (left) and per terrorist attack (right) happening in a given country.}
\label{fig:en_map}
\end{figure*}

\section{Limitations}
%Since we're applying our methods to terrorist attacks, if the database of terrorist attacks is incomplete, or biased, then the results of the analysis may not be representative of all terrorist attacks. 

In this study, we proposed minimalist fingerprints, which did not result in any fingerprint collisions for disasters, but resulted in 2\% of terrorist attacks having the same fingerprint. For other datasets, future studies may want to use longer fingerprints or account for the potential fingerprint collisions.

Our findings are inherently limited by the datasets we focus on. Terrorism is a somewhat subjective label, and an attack that is not considered a terrorist attack by the GTD will not register in our set. Similar caveats apply regarding EM-DAT: errors in the database will propagate to our analysis. Media Cloud, while far reaching, does not have universal coverage. 
% It is possible that Media Cloud, which has strong partnerships with African media organizations, might have better sub-Saharan African coverage than Latin American coverage. 
% This might, in turn, explain why we see French-language news sources paying close attention to disasters in Francophone Africa, but do not see a similar pattern in Latin America.

One serious limitation is the short duration of our study. Unlike \citeauthor{eisensee2007news}, who examine 34 years of data, we consider only six months. Those six months in question were quite unusual: COVID-19 was spreading rapidly, and our study period includes the global ``shutdown'' associated with the pandemic. It is possible that results during our study period might be significantly different than other time periods due to rapid changes in global newsgathering and the global focus on the pandemic. We are reassured though that our method seems to emphasize similarities in media attention to studies conducted during other time periods.

\section{Conclusion}

In this study, we present a novel methodology for systematically identifying and investigating global news coverage of critical events, specifically disasters and terrorist attacks. Our method, \textsc{FAME} (\textsc{Fingerprint to Article Matching for Events}), provides a scalable, efficient, and robust approach to match event fingerprints—time, location, and class—to news articles without relying on training data. By leveraging massive databases, including event databases and multilingual news articles from Media Cloud, \textsc{FAME} achieved excellent performance, with average F1 of $94.0$ on English articles, and of $95.3$ across all three languages. It linked 470 events to over 27,441 news articles (out of 42.4 million) across 3 languages and 8 event classes. The introduced novel dataset and task of linking events from databases to news articles, will stimulate the development of methods that will enable unprecedented agenda setting studies. Future works can extend our approach, e.g., by using more complex fingerprints, or supervised learning.

Through our large-scale analysis, we identified patterns in global news coverage that align with findings from prior research and add nuance to them. Specifically, we observed that events with higher death tolls, occurring in countries with higher GDP, and those with greater trade volumes are more likely to receive extensive global media attention (noticeable in Figure~\ref{fig:en_map}). Furthermore, this methodology enables us to investigate such patterns at scale, overcoming the limitations of prior approaches.

This work provides a robust foundation for further exploration of patterns in global news coverage. By offering a scalable and automated solution, \textsc{FAME} opens avenues for examining critical events across centuries and their coverage in diverse languages, regions, and contexts. 
Future research can build on this approach to deepen our understanding of how global media frames and prioritizes events.
% , shedding light on the socio-political and economic factors that influence media attention.

\section*{Acknowledgments}
We thank Paul Musgrave for extensive feedback and preliminary collaboration, and additionally the UMass NLP group and the anonymous reviewers for helpful comments and feedback. This work was supported by the Provost’s Interdisciplinary Research Grant 2023 (University of Massachusetts Amherst) titled ``Analyzing cross-country bias in news coverage of international conflicts and disasters'',
as well as the National Science Foundation award 1845576.
Any opinions, findings, and conclusions or recommendations expressed in
this material are those of the authors and do not necessarily reflect
the views of the National Science Foundation. 
For annotations, we thank Ana Lucia Hatsumi Yamashiro Miyahira, Luciano Redondo, Romaric Zoungan, Olivier Gaurat, and Houndemikon Hintekpo David.% and several others who chose to remain anonymous.

\bibliography{aaai22}

\section*{Paper Checklist}

\begin{enumerate}

\item For most authors...
\begin{enumerate}
    \item  Would answering this research question advance science without violating social contracts, such as violating privacy norms, perpetuating unfair profiling, exacerbating the socio-economic divide, or implying disrespect to societies or cultures? \textcolor{red}{Yes.}
  \item Do your main claims in the abstract and introduction accurately reflect the paper's contributions and scope?
    \textcolor{red}{Yes.}
   \item Do you clarify how the proposed methodological approach is appropriate for the claims made? 
    \textcolor{red}{Yes.}
   \item Do you clarify what are possible artifacts in the data used, given population-specific distributions?
    \textcolor{gray}{NA.}
  \item Did you describe the limitations of your work?
    \textcolor{red}{Yes.}
  \item Did you discuss any potential negative societal impacts of your work?
    \textcolor{blue}{No, because we haven't found any potential negative societal impacts.}
      \item Did you discuss any potential misuse of your work? \textcolor{gray}{NA.}
    \item Did you describe steps taken to prevent or mitigate potential negative outcomes of the research, such as data and model documentation, data anonymization, responsible release, access control, and the reproducibility of findings?
    \textcolor{red}{Yes.}
  \item Have you read the ethics review guidelines and ensured that your paper conforms to them?
    \textcolor{red}{Yes.}
\end{enumerate}

\item Additionally, if your study involves hypotheses testing...
\begin{enumerate}
  \item Did you clearly state the assumptions underlying all theoretical results?
    \textcolor{gray}{NA.}
  \item Have you provided justifications for all theoretical results?
    \textcolor{gray}{NA.}
  \item Did you discuss competing hypotheses or theories that might challenge or complement your theoretical results?
    \textcolor{gray}{NA.}
  \item Have you considered alternative mechanisms or explanations that might account for the same outcomes observed in your study?
    \textcolor{red}{Yes.}
  \item Did you address potential biases or limitations in your theoretical framework?
    \textcolor{gray}{NA.}
  \item Have you related your theoretical results to the existing literature in social science?
    \textcolor{red}{Yes.}
  \item Did you discuss the implications of your theoretical results for policy, practice, or further research in the social science domain?
    \textcolor{red}{Yes.}
\end{enumerate}

\item Additionally, if you are including theoretical proofs...
\begin{enumerate}
  \item Did you state the full set of assumptions of all theoretical results?
    \textcolor{gray}{NA.}
	\item Did you include complete proofs of all theoretical results?
    \textcolor{gray}{NA.}
\end{enumerate}

\item Additionally, if you ran machine learning experiments...
\begin{enumerate}
  \item Did you include the code, data, and instructions needed to reproduce the main experimental results (either in the supplemental material or as a URL)?
    \textcolor{red}{Yes.}
  \item Did you specify all the training details (e.g., data splits, hyperparameters, how they were chosen)?
    \textcolor{red}{Yes.}
     \item Did you report error bars (e.g., with respect to the random seed after running experiments multiple times)?
    \textcolor{red}{Yes.}
	\item Did you include the total amount of compute and the type of resources used (e.g., type of GPUs, internal cluster, or cloud provider)?
    \textcolor{gray}{NA.}
     \item Do you justify how the proposed evaluation is sufficient and appropriate to the claims made? 
    \textcolor{red}{Yes.}
     \item Do you discuss what is ``the cost`` of misclassification and fault (in)tolerance?
    \textcolor{red}{Yes.}
  
\end{enumerate}

\item Additionally, if you are using existing assets (e.g., code, data, models) or curating/releasing new assets, \textbf{without compromising anonymity}...
\begin{enumerate}
  \item If your work uses existing assets, did you cite the creators?
    \textcolor{red}{Yes.}
  \item Did you mention the license of the assets?
    \textcolor{gray}{NA.}
  \item Did you include any new assets in the supplemental material or as a URL?
    \textcolor{red}{Yes.}
  \item Did you discuss whether and how consent was obtained from people whose data you're using/curating?
    \textcolor{gray}{NA.}
  \item Did you discuss whether the data you are using/curating contains personally identifiable information or offensive content?
    \textcolor{red}{Yes.}
\item If you are curating or releasing new datasets, did you discuss how you intend to make your datasets FAIR (see \citet{fair})?
\textcolor{red}{Yes.}
\item If you are curating or releasing new datasets, did you create a Datasheet for the Dataset (see \citet{gebru2021datasheets})? 
\textcolor{red}{Yes.}
\end{enumerate}

\item Additionally, if you used crowdsourcing or conducted research with human subjects, \textbf{without compromising anonymity}...
\begin{enumerate}
  \item Did you include the full text of instructions given to participants and screenshots?
    \textcolor{red}{Yes.}
  \item Did you describe any potential participant risks, with mentions of Institutional Review Board (IRB) approvals?
    \textcolor{red}{Yes.}
  \item Did you include the estimated hourly wage paid to participants and the total amount spent on participant compensation?
    \textcolor{red}{Yes.}
   \item Did you discuss how data is stored, shared, and deidentified?
   \textcolor{red}{Yes.}
\end{enumerate}

\end{enumerate}

% \appendix

\appendix

\section{Appendix A: Per-event Evaluation for Spanish- and French-language articles}
\label{app:per-event-eval-es-fr}

\begin{figure}
    \centering
    \includegraphics[scale=0.33]{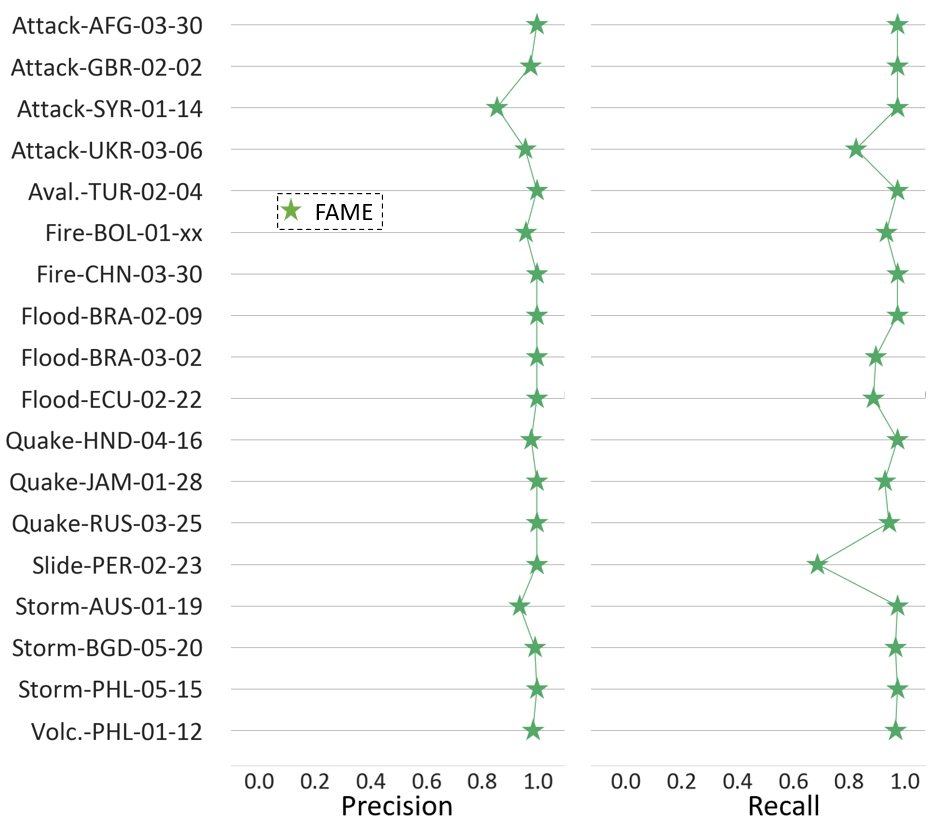}
    \caption{Per-event precision and recall of \textsc{FAME}, on linking Spanish-language articles to event fingerprints.}
    \label{fig:performance-es}
\end{figure}

\begin{figure}
    \centering
    \includegraphics[scale=0.33]{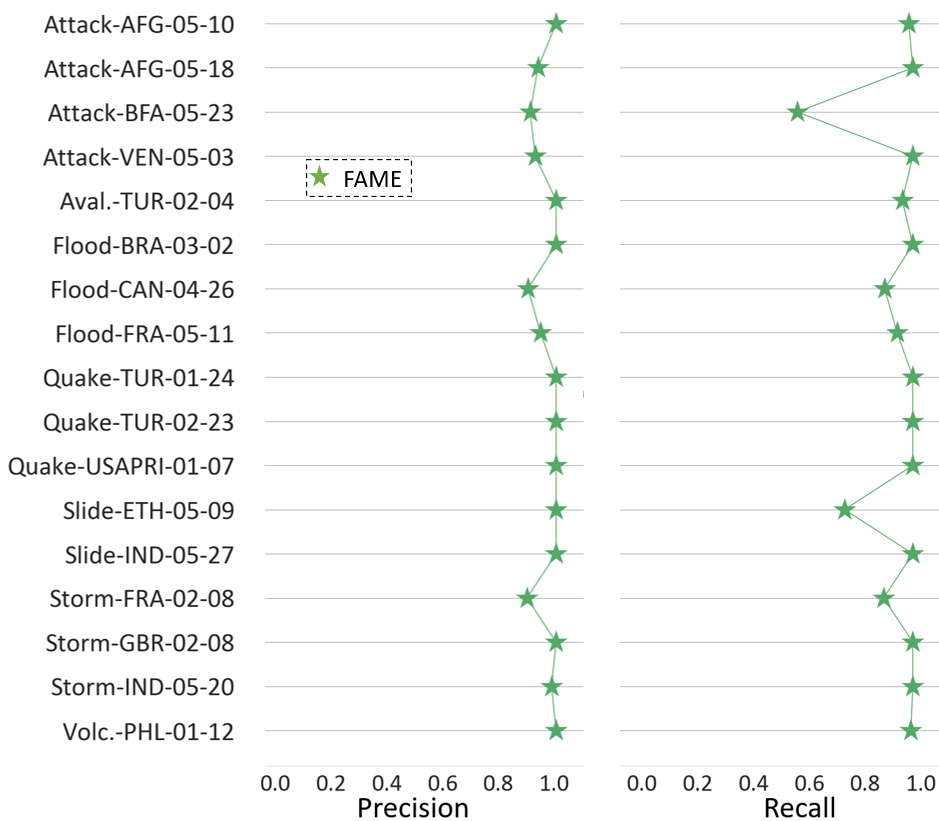}
    \caption{Per-event precision and recall of \textsc{FAME}, on linking French-language articles to event fingerprints.}
    \label{fig:performance-fr}
\end{figure}

Figures \ref{fig:performance-es} and \ref{fig:performance-fr} present the per-event precision and recall scores of \textsc{FAME}, on linking disaster events of $8$ classes with labeled Spanish- and French-language articles, where events must have at least one article discussing them to properly calculate precision and recall. In most cases, \textsc{FAME} can achieve more than 90\% score on the metrics. 

\section{Appendix B: Ablations}
\label{app:ablation}

%: we add further peripheral synonyms for event classes to event class keyword sets $K_c$, but find that the number of positive articles that discuss an event does not change. 

\noindent \textbf{Increasing the size of $K_c$ and $K_{l}$.} We investigate whether the keyword sets that \textsc{FAME} uses are comprehensive enough. We further increase the size of $K_c$ by including $5$ to $10$ more synonyms and hyponyms that GPT-3.5-turbo recommends for each event class $c$, even if they seem to have a very different meaning from $c$, but observed no difference in results compared to when using the original keyword sets. %\bto{how much do the sets change? for example, what are the sizes before and after this expansion?} 
%We observed no difference in results using the original keyword sets. compared to the more flexible keyword sets. 

\noindent \textbf{Using open-source LLMs.} We replace GPT-3.5-turbo in the Phase Two LLM QA step with recent open-source models. In Table~\ref{tab:performance-with-opensourcellms}, we show results of \textsc{FAME} using high-performing open-source models of llama-3-70b and llama-3-8b, which perform quite similarly to, but slightly worse, than \textsc{FAME} using GPT-3.5-turbo. The difference in number of model parameters does not seem to impact performance much.

\begin{table}{\footnotesize
\begin{center}
\begin{tabular}{ccccc}\toprule
\textbf{Lang.}& \textbf{Method} & \textbf{Prec.} & \textbf{Recall} & \textbf{F1} \\ \midrule%\cmidrule(lr)[4-4]\cmidrule(lr)[5-5]
EN & KW: Title-only & 88.1 & 48.4 & 57.6 \\
EN & KW: Title+Body & 84.1 & 78.0 & 79.3 \\
\midrule
EN &	\textsc{FAME} (llama-3-70b)	& 89.0 & 84.3 & 85.5\\
EN &	\textsc{FAME} (llama-3-8b)	& 90.9 & 86.1 & 87.0\\
EN & \textsc{FAME} (GPT-3.5-turbo) & 93.2 & 95.4 & 94.0 \\
\midrule
FR & \textsc{FAME} (llama-3-70b) & 97.6 & 90.5 & 93.3\\
FR & \textsc{FAME} (llama-3-8b) & 94.8 & 93.9 & 94.0\\
FR & \textsc{FAME} (GPT-3.5-turbo) & 96.9 & 94.1 & 95.2 \\
\midrule
ES & \textsc{FAME} (llama-3-70b) & 98.2 & 91.4 & 94.2\\
ES & \textsc{FAME} (llama-3-8b) & 96.8 & 93.0 & 94.5\\
ES & \textsc{FAME} (GPT-3.5-turbo) & 98.1 & 96.0 & 96.8 \\
\bottomrule
\end{tabular}\end{center}\caption{Averaged precision, recall, and F1 scores over expert-annotated articles corresponding to events that do not have zero prevalence, for English (\textsc{FAME} and baselines), Spanish (\textsc{FAME}), and French (\textsc{FAME}). }
\label{tab:performance-with-opensourcellms}}

\end{table}

\noindent \textbf{Question specificity.} We also investigate whether wording of the LLM prompt affects performance. %The first wording asks if the event class has occurred in a country recently. The next two prompts clarify the broader category of the event class (e.g., natural disaster), using slightly different wordings. The fourth prompt provides the definition of the event class. 
First, we find that using a single prompt for the Phase Two LLM QA step has better performance than using a series of prompts. We therefore experiment with various single prompts:

\vspace{.3em}
\fbox{\begin{minipage}{21em}\texttt{[article title + first 3 sentences]}\\

\noindent\texttt{Does the text discuss a recent [event class $c$] in [location $l$]?}
\end{minipage}}
\vspace{.5em}

\noindent This wording is the simplest, asking if the event class in the fingerprint has occurred in the country specified by the fingerprint recently. 

\vspace{.3em}
\fbox{\begin{minipage}{21em}\texttt{[article title + first 3 sentences]}\\

\noindent\texttt{Does the text discuss a recent [event class $c$] [broader category of $c$] in [location $l$]?}
\end{minipage}}
\vspace{.5em}

\vspace{.3em}
\fbox{\begin{minipage}{21em}\texttt{[article title + first 3 sentences]}\\

\noindent\texttt{Does the text discuss a recent [event class $c$] [(broader category of $c$)] in [location $l$]?}
\end{minipage}}
\vspace{.5em}

\noindent These two prompts clarify the broader category of the event class (e.g., natural disaster, terrorist attack), using slightly different wordings.

\vspace{.3em}
\fbox{\begin{minipage}{21em}\texttt{[article title + first 3 sentences]}\\

\noindent\texttt{Does the text discuss a recent [event class $c$], where [definition of $c$], in [location $l$]?}
\end{minipage}}
\vspace{.5em}

\noindent The fourth prompt provides the definition of the event class, where the definition is from the Merriam-Webster dictionary. 

The results corresponding to various prompts are all within one percentage point of each other for precision, recall, and F1 score. To avoid prompt customization for each event class, \textsc{FAME}'s Phase Two LLM QA step uses the simplest prompt without any category or definition clarification.  %Thus, we selected the question discussed in (Appendix).

%The first wording asks if the event class has occurred in a country recently. The next two prompts clarify the broader category of the event class (e.g., natural disaster), using slightly different wordings. The fourth prompt provides the definition of the event class. Results...

\section{Appendix C: Reporting countries and example of identified terrorist attacks}
\label{app:reporting-countries-examples-of-attacks}

\begin{table}[htb]{\small
\caption{Reporting Countries by Official Language}
\label{tab:report_countries_by_language}
\begin{center}
\begin{tabular}{ccc}
\toprule
\textbf{English (en)} & \textbf{Spanish (es)} & \textbf{French (fr)} \\
\midrule
United States (USA)   & Mexico (MEX)          & Canada (CAN) \\
United Kingdom (GBR)  & Colombia (COL)        & France (FRA) \\
Canada (CAN)          & Spain (ESP)           & \\
Australia (AUS)       & Argentina (ARG)       & \\
Ghana (GHA)           &                       & \\
India (IND)           &                       & \\
Ireland (IRL)         &                       & \\
Kenya (KEN)           &                       & \\
Nigeria (NGA)         &                       & \\
South Africa (ZAF)    &                       & \\
\bottomrule
\end{tabular}
\end{center}
}
\end{table}

Table \ref{tab:report_countries_by_language} presents the list of reporting countries we considered in the regression on global news coverage.

Similar to Table \ref{tab:top_disasters_2020}, we also examined example terrorist attacks identified by our method as part of a quantitative robustness check. Table \ref{tab:top_conflicts_2020} lists the top 10 terrorist attacks that garnered the highest number of news reports. Most of these identified terrorist attacks align with online news records, except for one false positive case. Specifically, our database includes two attack events, "Woman shot in Temple Terrace" and "Boogaloo shooting in Oakland," which overlap with the period of the "George Floyd protests" series of events. All three events share the same fingerprint. As a result, our method erroneously categorized thousands of news articles on the "George Floyd protests" as false positive samples associated with the fingerprint of an attack event.

\begin{table*}[ht!]
\centering
\begin{tabular}{l|lrl|lll}
\toprule
\textbf{Event} & \textbf{Class} ($c$) & \textbf{Location} ($l$) & \textbf{Time} ($t$) & \textbf{News} & \textbf{Deaths} \\
\midrule
Woman shot in Temple Terrace & \multirow{2}{*}{Attack}  & \multirow{2}{*}{United States} & \multirow{2}{*}{2020-05-29} & \multirow{2}{*}{15094} & 1\\
Boogaloo shooting in Oakland & & & & & 1 \\
% https://en.wikipedia.org/wiki/2020_boogaloo_murders}{Boogaloo shooting in Oakland} & Attack & United States & 2020-05-29 & 15094 & 1\\
Foiled coup attempt & Attack & Venezuela & 2020-05-03 & 1741 & 6\\
Shooting at Corpus Christi Navy base & Attack & United States & 2020-05-21 & 1712 & 1 \\
Hanau shootings & Attack & Germany & 2020-02-19 & 1677 & 2 \\
Fulani militants attacked in Kaduna State & Attack & Nigeria & 2020-05-31 & 1379 & 13 \\
Massacre in a maternity ward & Attack & Afghanistan & 2020-05-12 & 1216 & 24 \\
Streatham stabbing & Attack & United Kingdom & 2020-02-02 & 1161 & 1 \\
Gamboru bombing & Attack & Nigeria & 2020-01-06 & 1024 & 30 \\
Jorge Armenta: Mexican journalist killed & Attack & Mexico & 2020-05-16 & 1022 & 2\\
Boko Haram militants burn to death motorists & Attack & Nigeria & 2020-02-09 & 980 & 30 \\
\bottomrule
\end{tabular}
\caption{Top 10 terrorist attacks with the most news in the first five months of 2020.}
\label{tab:top_conflicts_2020}
\end{table*}

% \begin{table*}[h!]
% \centering
% \begin{tabular}{l|lrl|lll}
% \toprule
% \textbf{Event} & \textbf{Class} ($c$) & \textbf{Location} ($l$) & \textbf{Time} ($t$) & \textbf{News} & \textbf{Deaths} \\
% \midrule
% \href{https://www.tampabay.com/news/breaking-news/2020/05/29/officer-shoots-kills-woman-at-temple-terrace-city-hall-after-she-charged-with-knife-police-say/}{Woman shot in Temple Terrace} & \multirow{2}{*}{Attack}  & \multirow{2}{*}{United States} & \multirow{2}{*}{2020-05-29} & \multirow{2}{*}{15094} & 1\\
% \href{https://en.wikipedia.org/wiki/2020_boogaloo_murders}{Boogaloo shooting in Oakland} & & & & & 1 \\
% % \href{https://en.wikipedia.org/wiki/2020_boogaloo_murders}{Boogaloo shooting in Oakland} & Attack & United States & 2020-05-29 & 15094 & 1\\
% \href{https://www.nytimes.com/2020/05/03/world/americas/venezuela-coup.html}{Foiled coup attempt} & Attack & Venezuela & 2020-05-03 & 1741 & 6\\
% \href{https://www.washingtonpost.com/national-security/shooting-at-corpus-christi-navy-base-investigated-as-terrorist-attack-fbi-says/2020/05/21/63111df2-9ba8-11ea-ac72-3841fcc9b35f_story.html}{Shooting at Corpus Christi Navy base} & Attack & United States & 2020-05-21 & 1712 & 1 \\
% \href{https://en.wikipedia.org/wiki/Hanau_shootings}{Hanau shootings} & Attack & Germany & 2020-02-19 & 1677 & 2 \\
% \href{https://www.persecution.org/2020/05/31/fulani-militants-attacked-christians-kaduna-state-twenty-killed/}{Fulani militants attacked in Kaduna State} & Attack & Nigeria & 2020-05-31 & 1379 & 13 \\
% \href{https://www.doctorswithoutborders.org/latest/afghanistan-massacre-maternity-ward}{Massacre in a maternity ward} & Attack & Afghanistan & 2020-05-12 & 1216 & 24 \\
% \href{https://en.wikipedia.org/wiki/2020_Streatham_stabbing#:~:text=On%202%20February%202020%2C%20two,broken%20glass%20as%20a%20result.}{Streatham stabbing} & Attack & United Kingdom & 2020-02-02 & 1161 & 1 \\
% \href{https://en.wikipedia.org/wiki/2020_Gamboru_bombing}{Gamboru bombing} & Attack & Nigeria & 2020-01-06 & 1024 & 30 \\
% \href{https://www.bbc.com/news/world-latin-america-52698127}{Jorge Armenta: Mexican journalist killed} & Attack & Mexico & 2020-05-16 & 1022 & 2\\
% \href{https://www.bbc.com/news/world-africa-51445070}{Boko Haram militants burn to death motorists} & Attack & Nigeria & 2020-02-09 & 980 & 30 \\
% \bottomrule
% \end{tabular}
% \caption{Top 10 terrorist attacks with the most news in the first five months of 2020.}
% \label{tab:top_conflicts_2020}
% \end{table*}

\section{Appendix D: Country factors in regression analysis}
\label{app:country-factors-regression}

We extensively examined literature of the country factors whose impact on media coverage has been studied. All considered factors are listed (color-coded by the class) with detailed explanation in Table \ref{tab:country_factors}.

\begin{table*}[t!]{
% \begin{table}[t!]
\vspace{-0.3cm}
    % \NoHyper
    %\rowcolors{2}{gray!10}{white}
    \begin{adjustbox}{width = 2\columnwidth}
    \begin{tabular}{r r  p{1.8\columnwidth} r}
    \\ %Inserting one extra line because of a bad break with the word "Each"
    \toprule
    Type & Predictor & description & Threshold/range \\
    \toprule
    Eve. & \text{{\color{blue} \#Deaths}} & The average death number of each event happening in the country & [0, 284]\\
    \hline
   \multirow{3}{*}{Eco.} 
            & \text{{\color{orange} GDP}} & event in a high GDP country? & \multirow{5}{*}{GDP$>\$500$} \\
            & \text{{\color{orange} GDP (h-h)}}   & event in a high GDP country and reported by a high GDP country? & \\
            & \text{{\color{orange} GDP (h-l)}}   & event in a high GDP country and reported by a low GDP country? &  \\
            & \text{{\color{orange} GDP (l-h)}}   & event in a low GDP country and reported by a high GDP country? &  \\
            & \text{{\color{orange} GDP (l-l)}}   & event in a low GDP country and reported by a low GDP country? & \\
            & \text{{\color{orange} Trade}}  & the trading volume between the reporting country and the event country & [0, 740B]\\    
            & \text{{\color{orange} Investment}} &the amount (\$ USD) that the reporting country invest into the event country & [0, 112B]\\
    \hline
    \multirow{18}{*}{Pol.} & \text{{\color{teal} Democracy in.}} & event in a high democracy index (DI) country?   & \multirow{5}{*}{DI$>5$}\\
                            & \text{{\color{teal} Democracy in. (h-h)}}   & event in a high democracy index country and reported by a high democracy index country? &  \\
                        & \text{{\color{teal} Democracy in. (h-l)}}   & event in a high democracy index country and reported by a low democracy index country? &  \\
                        & \text{{\color{teal} Democracy in. (l-h)}}   & event in a low democracy index country and reported by a high democracy index country? &  \\
                        & \text{{\color{teal} Democracy in. (l-l)}}   & event in a low democracy index country and reported by a low democracy index country? &  \\
                         % & \text{{\color{teal} NATO}} & binary factor to classify if the event country is a NATO member & \{0, 1\} \\
                         % & \text{{\color{teal} EU}} & binary factor to classify if the event country is a EU member & \{0, 1\} \\
                         % & \text{{\color{teal} BRICS}} & binary factor to classify if the event country is a BRICS member & \{0, 1\} \\
                         % & \text{{\color{teal} both NATO}} & binary factor to classify if both the event country and the reporting country are NATO members  & \{0, 1\} \\
                         % & \text{{\color{teal} both EU}} & binary factor to classify if both the event country and the reporting country are EU members & \{0, 1\} \\
                         % & \text{{\color{teal} both BRICS}} & binary factor to classify if both the event country and the reporting country are BRICS members & \{0, 1\} \\
                         % & \text{{\color{teal} Across NATO–EU groups}} & binary factor to classify if the event country is a NATO member and the reporting country is a EU member  & \{0, 1\} \\
                         % & \text{{\color{teal} Across NATO–BRICS groups}} & binary factor to classify if the event country is a NATO member and the reporting country is a BRICS member & \{0, 1\} \\
                         % & \text{{\color{teal} Across EU–BRICS groups}} & binary factor to classify if the event country is a EU member and the reporting country is a BRICS member & \{0, 1\} \\
                         
                         & \text{{\color{teal} Press Freedom in.}} & event in a high press freedom index (PFI) country?   & \multirow{5}{*}{PFI$>50$}\\
                            & \text{{\color{teal} Press freedom in. (h-h)}}   & event in a high press freedom index country and reported by a high press freedom index country? &  \\
                        & \text{{\color{teal} Press freedom in. (h-l)}}   & event in a high press freedom index country and reported by a low press freedom index country? &  \\
                        & \text{{\color{teal} Press freedom in. (l-h)}}   & event in a low press freedom index country and reported by a high press freedom index country? &  \\
                        & \text{{\color{teal} Press freedom in. (l-l)}}   & event in a low press freedom index country and reported by a low press freedom index country? &  \\
                         
                         &  \text{{\color{teal} Federalism}}  & event in a federal country? & \{0, 1\}\\ 
                         &  \text{{\color{teal} Republic}}  & event in a republic country? & \{0, 1\}\\ 
                         &  \text{{\color{teal} Other mode of government}}  & event in a country neither federal or republic? & \{0, 1\}\\ 
                         &  \text{{\color{teal} Both Federalism}}  & both the event country and the reporting country are federal? & \{0, 1\}\\ 
                         &  \text{{\color{teal} Both Republic}}  & both the event country and the reporting country are republic? & \{0, 1\}\\  
                         &  \text{{\color{teal} Federalism and Other}} & event in a federal country and reported by a country neither federal or republic? & \{0, 1\}\\
                         &  \text{{\color{teal} Republic and Other}}   & event in a republic country and reported by a country neither federal or republic? & \{0, 1\}\\
                         
                         & \text{{\color{teal} Diplomatic relation}} & an integer measure with six categories: 1-Unknown; 2-Interests Served by; 3-Interest Desk; 4-Charge d’affairs; 5-Minister/Envoy;  6-Ambassador, Nuncio, Secretary of the People’s Bureau; & \{1, 2, 3, 4, 5, 6\} \\
    \hline
    \multirow{8}{*}{Soc.} &  \text{{\color{brown} Population}}  & event country population & [2.5M, 1.4B]\\
                          & \text{{\color{brown} Population density}} & the population density of the event country & \\
                        & \text{{\color{brown} Immigration}}   & the immigration amount from the reporting country to the event country & [0, 11.6M]\\

& \text{{\color{brown} Gini in.}} & event in a high Gini index country?   & \multirow{5}{*}{Gini$>50$}\\
                            & \text{{\color{brown} Gini in. (h-h)}}   & event in a high Gini index country and reported by a high Gini index country? &  \\
                        & \text{{\color{brown} Gini in. (h-l)}}   & event in a high Gini index country and reported by a low Gini index country? &  \\
                        & \text{{\color{brown} Gini in. (l-h)}}   & event in a low Gini index country and reported by a high Gini index country? &  \\
                        & \text{{\color{brown} Gini in. (l-l)}}   & event in a low Gini index country and reported by a low Gini index country? &  \\
    \hline
    \multirow{4}{*}{Geo.}  & \text{{\color{Green} Area}} & the area of the event country & [21K, 10M] \\
                           & \text{{\color{Green} Asia}} & event country in Asia? & \{0, 1\}\\
                          & \text{{\color{Green} Europe}} & event country in Europe? & \{0, 1\}\\
                          & \text{{\color{Green} Africa}} & event country in Africa? & \{0, 1\}\\
                          & \text{{\color{Green} Oceania}} & event country in Oceania? & \{0, 1\}\\
                          & \text{{\color{Green} North America}} &  event country in North America? & \{0, 1\}\\
                          & \text{{\color{Green} South America}} & event country in South America? & \{0, 1\}\\
                          & \text{{\color{Green} Neighbors}} & event country and the reporting country are neighbors? & \{0, 1\}\\
                          & \text{{\color{Green} Continent sim}} & event country and the reporting country on the same continent? & \{0, 1\}\\
    \hline
    \multirow{5}{*}{Cul.} & \text{{\color{purple} Same Language}} &  the event country and the reporting country using the same official language? & \{0, 1\} \\
                        & \text{{\color{purple} Religion Diversity}} & the shannon entropy of the distribution with respect to mainstream religions (including Christian, Jewish, Muslim, Unaffil, Hindu, Buddhist, and Folk religion) in the event country & [0.40, 0.97]\\
                        % & \text{{\color{purple} Media outlets number}}  & the number of media outlets in the reporting country \\
                        & \text{{\color{purple} Literacy rate}} & the literacy rate of the reporting country & [0.49, 1] \\ 
                        & \text{{\color{purple} Internet user rate}} & the internet user rate of the reporting country & [0.13, 0.92] \\
    \bottomrule
    \end{tabular}
    \end{adjustbox}
    \caption{Descriptions for the predictors considered in the regressions of news coverage.}
    \label{tab:country_factors}
}

\end{table*}

\begin{figure*}[t!]
\centering
\includegraphics[width=0.98\columnwidth]{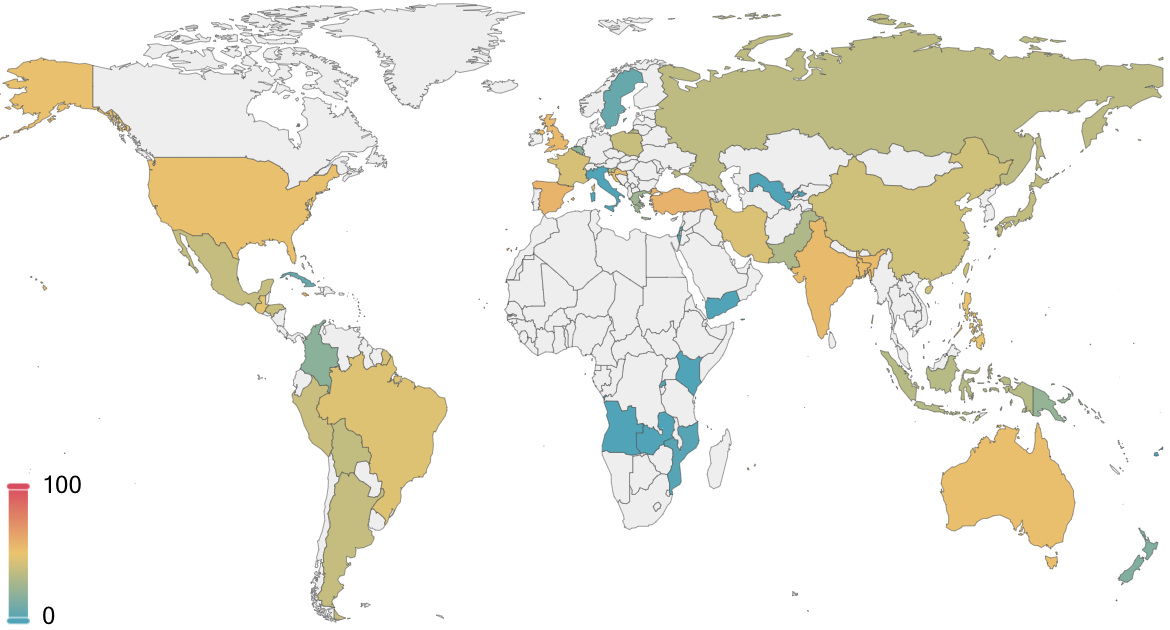}
\includegraphics[width=0.98\columnwidth]{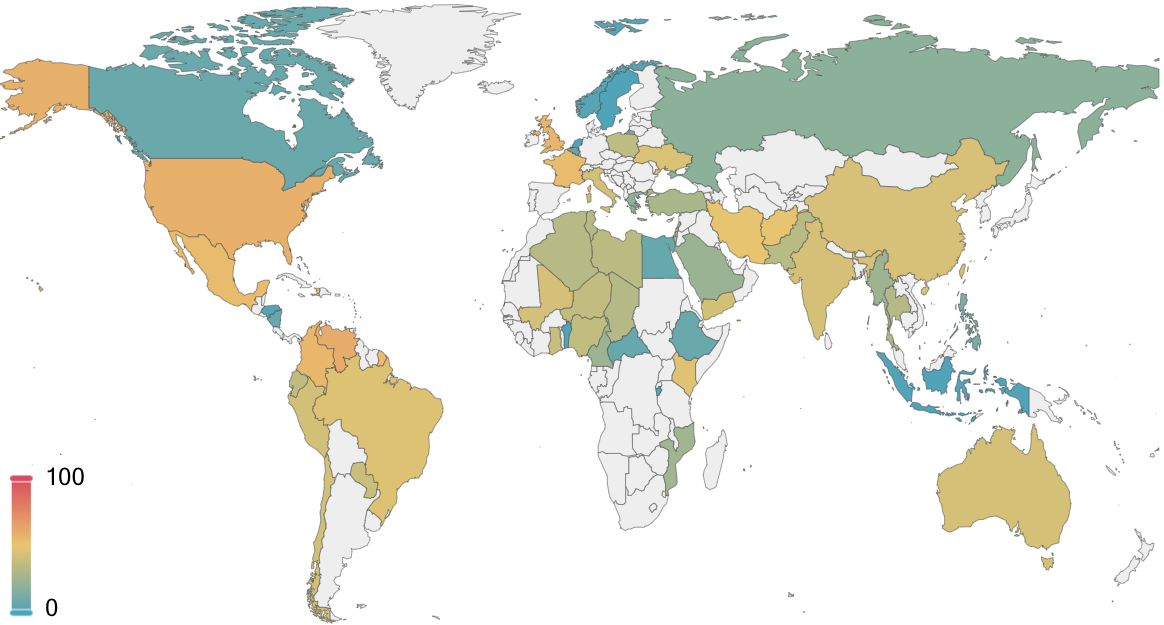}
\caption{The average number of news in Spanish per disaster (left) and per terrorist attack (right) happening in a given country.
% Interestingly an attack in Ukraine is reported twice of one in Israel.
}
\label{fig:es_map}
\end{figure*}

\begin{figure*}[t!]
\centering
\includegraphics[width=0.98\columnwidth]{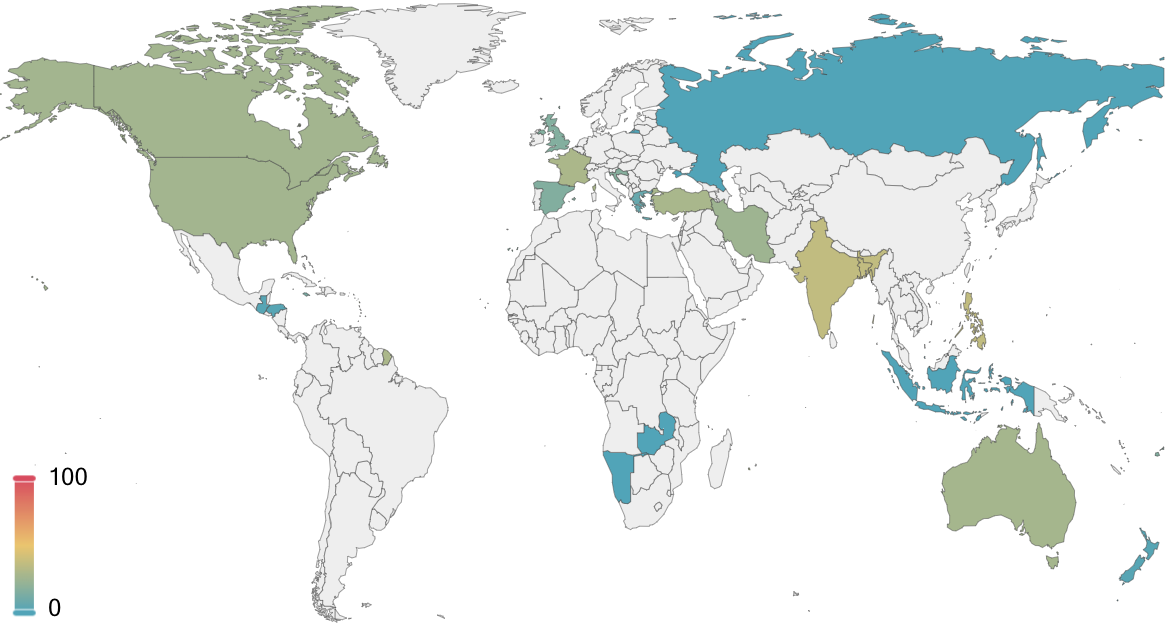}
\includegraphics[width=0.98\columnwidth]{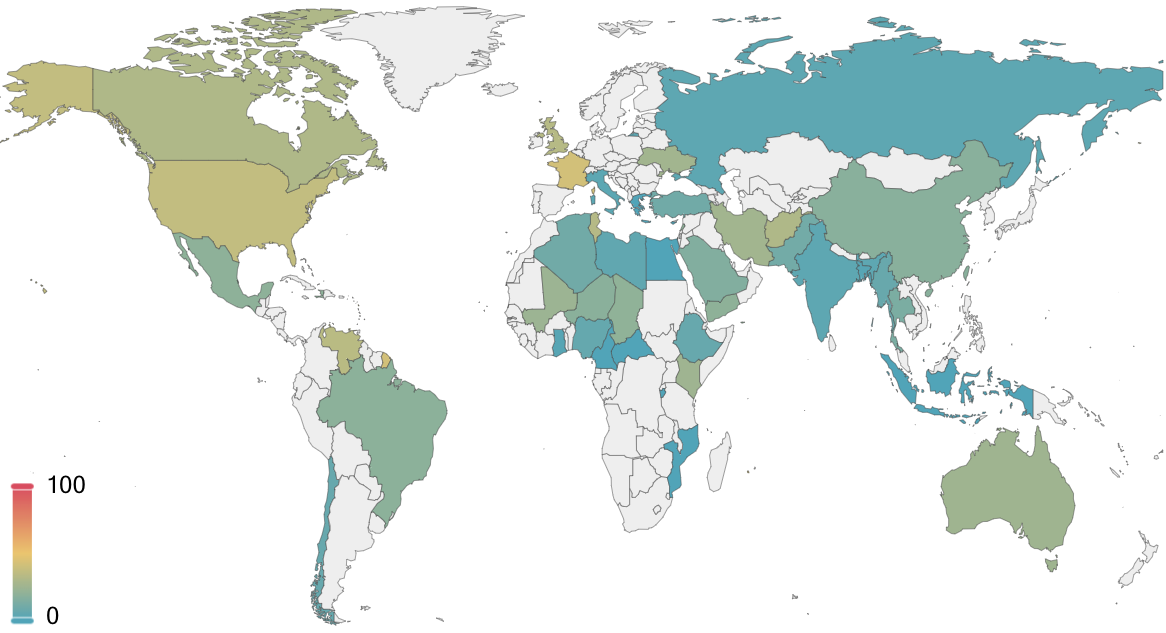}
\caption{The average number of news in French per disaster (left) and per terrorist attack (right) happening in a given country.}
\label{fig:fr_map}
\end{figure*}

\end{document}